\documentclass{iopart}
\usepackage{xspace}

\usepackage{graphicx,amsfonts,amssymb,amsmath,hyperref}
\usepackage{textcomp}

\newif\ifhyper
\hypertrue
\ifhyper
\hypersetup{
  citecolor = {green},
  colorlinks = {true}, 
  urlcolor = {blue} 
}

\newcommand{\p}{\partial} 

\newcommand{\tp}{\tilde{\psi}}

\newcommand{\tphi}{\tilde{\phi}}

\newcommand{\btheta}{\bar{\theta}}

\newcommand{\tJ}{\tilde{j}}

\newcommand{\bD}{\bar{D}}
\newcommand{\bQ}{\bar{Q}}
\newcommand{\bZ}{\bar{Z}}
\newcommand{\bX}{\bar{X}}
\newcommand{\coefu}{\Omega_1}
\newcommand{\coefd}{\Omega_2}
\newcommand{\coef}{}
\newcommand{\hr}{\hat{\psi}}
\newcommand{\hu}{\hat{u}}
\newcommand{\hz}{{\hat{z}}}
\newcommand{\hx}{\hat{x}}
\newcommand{\vx}{{\vec x}}
\newcommand{\vq}{{\vec q}}
\newcommand{\cD}{{\cal D}}
\newcommand{\la}{\langle}
\newcommand{\ra}{\rangle}
\newcommand{\etaz}{\eta_{\hbox{\tiny Z}}}
\newcommand{\etax}{\eta_{\hbox{\tiny X}}}

\newcommand{\anz}{{\em Ansatz}\xspace}
\newcommand{\nequ}{non-equilibrium\xspace}

\begin{document}

\title[ General framework of NPRG for NESS]{General framework of the non-perturbative 
renormalization group for  non-equilibrium steady states}
\author{L\'eonie Canet}
\address{ Laboratoire de Physique et
 Mod\'elisation des Milieux Condens\'es, Universit\'e Joseph Fourier Grenoble I -- CNRS, BP166,  38042 Grenoble Cedex, France}

\author{Hugues Chat\'e}
\address{Service de Physique de l'\'Etat Condens\'e, CEA - Saclay, 91191
~Gif-sur-Yvette Cedex,~France}
\author{Bertrand Delamotte}

\address{Laboratoire de Physique Th\'eorique de la Mati\`ere Condens\'ee, Universit\'e 
Pierre et Marie Curie - Paris VI, CNRS UMR 7600, 4 Place Jussieu, 75252 Paris Cedex 05, France}


\begin{abstract}   This   paper  is   devoted  to presenting in detail
 the non-perturbative  renormalization group
(NPRG)  formalism to  investigate out-of-equilibrium systems and  critical dynamics
 in statistical physics. The general NPRG framework for studying \nequ
 steady states in stochastic models is expounded  and  fundamental technicalities  are stressed,
mainly regarding the role of causality and of It$\bar{\rm o}$'s discretization.
We analyze the consequences  of It$\bar{\rm o}$'s  prescription  in the 
NPRG framework and eventually provide  an adequate
regularization to encode them automatically.
 Besides, we show how to build a supersymmetric NPRG formalism with
emphasis on time-reversal symmetric problems, whose supersymmetric
structure allows for a particularly simple implementation of NPRG in which
causality issues are transparent. We illustrate the two approaches on the example of Model A
within the  derivative expansion approximation at order two, and check that they
yield identical results.
We stress, though, that the framework presented here  also applies to genuinely
out-of-equilibrium problems.
\end{abstract}

\maketitle

\section{Introduction}
\label{intro}

The  theoretical   understanding  of  out-of-equilibrium   systems, and in particular non-equilibrium steady states, is
nowadays one of the major quests of statistical physics. These systems
exhibit a  great variety of behavior,  
including genuinely non-equilibrium phase
transitions and critical phenomena, ubiquitous strong-coupling regimes,
recurrent absence of lower critical dimension, etc. 
Numerous partial results have been obtained at various levels, but all-purpose, 
analytical, and systematic methods to treat non-equilibrium problems are scarce.

The   nonperturbative  renormalization   group  (NPRG) is one such method,
and this  paper is devoted to detail its implementation for out-of-equilibrium systems.  
Indeed, many remarkable results have already been obtained in statistical physics
using the NPRG approach not only at equilibrium in
systems such as frustrated magnets \cite{delamotte03}, the random field or random bond Ising 
model \cite{Tarjus2004},
membranes \cite{Essafi2011}, bosonic systems \cite{Dupuis2009}, 
but also out-of-equilibrium, where one can mention
important advances in reaction-diffusion systems \cite{canet04},
and for the Kardar-Parisi-Zhang universality class \cite{canet10}.

The starting point for applying NPRG methods is a field theory. 
Such a field theory is usually obtained by following one of two possible 
routes. Either one starts from an effective mesoscopic description in terms of  a
Langevin equation and  one then resorts to the Janssen-de Dominicis
formalism   \cite{janssen76}  to   construct  a   functional  integral
representation  of  this  equation,  upon  introducing  an  additional
Martin-Siggia-Rose response field  \cite{martin73}. Or,  
and this is the other route, one starts from  a
microscopic master equation that can be mapped onto a functional integral
following   the  Doi-Peliti  procedure   \cite{doi76}.   
When  both descriptions  coexist  the  resulting   field  theories  are  
of  course equivalent (see however \cite{benitez11} for important subtleties 
in the derivation of the equivalence of the Janssen-de Dominicis
and  Doi-Peliti field theories). 

Both field theoretical formulations are derived in discrete time, and
a crucial ambiguity arises when performing the continuous-time limit
\cite{janssen92,zinn}.   This  problem   implies  the   choice   of  a
prescription in the continuum version 
(usually related to the It$\bar{\rm o}$'s or Stratonovitch's discretization scheme),
 the  consequences of which  still deserve  discussions  ({\it e.g.  }
\cite{honkonen11}).  Roughly  speaking,  the  Janssen-de  Dominicis
transformation for instance involves  a determinant which depends on
the  original   discretization  of  time.  This   determinant  can  be
conveniently set  to unity upon adopting  It$\bar{\rm o}$'s discretization scheme and
imposing  the related  prescription in  subsequent  calculations. This
prescription  is   easily  implemented  in   perturbation  theory: it
amounts to cancel out tadpole diagrams \cite{zinn}. 

The  implications  of this choice of time-discretization within the  
NPRG formalism have yet to be clarified and is one of the main subjects 
of this paper. We shall first identify  the
consequences of It$\bar{\rm o}$'s discretization in NPRG flow equations and
then  device a simple procedure to systematically encode them.  
This will be the opportunity to review some other important technical aspects
arising when implementing, out-of-equilibrium, 
the most popular approximation scheme of  the exact NPRG  equation, 
the  derivative expansion \cite{berges02,canet03a}.

To guide us in  the  task of understanding the consequences of 
It$\bar{\rm o}$'s discretization in the NPRG framework,  
we first review, in section \ref{disc},
the discrete-time version of \nequ field theories ensuing from the two standard 
Janssen-de Dominicis and  Doi-Peliti procedures, analyze causality issues, and
recall the usual prescription to encode It$\bar{\rm o}$'s discretization in perturbation theory.
Section \ref{nprg} is dedicated to expounding
 the specificities of the NPRG framework when applied to \nequ systems. Causality   
issues are postponed to Section \ref{ito}, which focusses on identifying the relevant  prescription
associated with It$\bar{\rm o}$'s time-discretization in the NPRG context, and where
we propose a simple ``regularization'' procedure to encode it. 
Even though it can be (and has been) applied to genuinely out-of-equilibrium problems,
we illustrate it, in Section~\ref{modelA}, to a relaxation-towards-equilibrium problem, the critical dynamics of Model A.
In Section \ref{supersymmetry}, we show that for such models satisfying time-reversal symmetry, one can build a supersymmetric formulation of NPRG.
Not only this formulation is elegant  and  convenient for  actual  calculations but,  more
importantly here, it is free from any ambiguity regarding the continuous-time
limit \cite{zinn}. Once the supersymmetric NPRG formalism is set up, it is  applied to Model A, which allows us to check that the two approaches
yield identical results.

\section{ Field theories for nonequilibrium systems}
\label{disc}

As it constitutes the basis of the following discussions,
 we start by recalling the derivation of the field theory associated with
an out-of-equilibrium stochastic model. We review  the two standard procedures,
 starting from a Langevin equation or starting from  a master equation. 
In both cases, the choice made for the 
continuous-time limit plays an important technical role both perturbatively and nonperturbatively.

\subsection{Doi-Peliti functional}

Many models (such as reaction-diffusion systems) are defined at a 
microscopic scale by a set of dynamical rules.  The time evolution of these systems
 is governed by a master equation, which, in an operator representation, can be written as
\begin{equation}
 \p_t |P(t)\ra = \hat{\cal W}  |P(t)\ra  
\label{master}
\end{equation}
where $|P(t)\ra = \sum_n P(n,t) | n\ra$ is the ket associated with the 
probability law of the system, $|n\ra$ are the Fock states (number states)   
and $\hat{\cal W} \equiv \hat{\cal W}(a^\dagger,a)$ is the operator, 
expressed in terms of creation and annihilation operators, representing the kernel of the master 
equation. Eq. (\ref{master}) can be cast into a field theory following the Doi-Peliti procedure \cite{doi76}. We
 do not review it in detail here (see {\it e.g.} \cite{doi76,cardy98}), but  merely stress a few points regarding the ordering of discrete times.

The mean value of an observable ${\cal O}$ at time $t$ can be expressed in the operator representation 
 as 
\begin{equation}
 \la {\cal O}\ra(t) = \la {\cal P} |\hat {\cal O}e^{\hat{\cal W}t}| P(0)\ra
\end{equation}
where $\la {\cal P}|$ is the projection state satisfying $\la {\cal P} |P(t)\ra=1$ and $\la {\cal P} |\hat{\cal W} =0$. 
The procedure consists in splitting the evolution operator $\exp(\hat{\cal W}t)$ in $N+1$ time slices of width $\tau$
and in inserting in each a closure relation written in terms of the coherent states $\vert\phi_i\rangle$:
\begin{equation}
 \la {\cal O} \ra(t) = \displaystyle{\int \left(\prod_{\alpha=0}^N\,  \frac{d^2\phi_\alpha}{\pi}\right) \la {\cal P} |\hat {\cal O}| 
\phi_N\ra \la\phi_{N} |e^{\hat{\cal W}\tau}|\phi_{N-1}\ra\dots \la\phi_{1} |e^{\hat{\cal W}\tau}|\phi_{0}\ra\la \phi_{0}|P(0)\ra}
\end{equation}
where $d^2\phi_\alpha \equiv d{\cal I}m(\phi_\alpha) d{\cal R}e(\phi_\alpha)$. At first order in $\tau$ and after re-exponentiation, it writes 
\begin{equation}
 \la {\cal O}\ra(t) = \displaystyle\int \left(\prod_{\alpha=0}^N\, \frac{d^2\phi_\alpha}{\pi}\right) \la {\cal P} |\hat {\cal O}| 
\phi_N\ra\prod_{k=1}^N \la\phi_k|\phi_{k-1}\ra \exp\left(\tau \sum_{k=1}^N 
\frac{ \la\phi_k|\hat{\cal W}|\phi_{k-1}\ra}{ \la\phi_k|\phi_{k-1}\ra}\right)\la\phi_{0}|P(0)\ra.
\end{equation}
By computing the usual nonvanishing overlap 
$\la\phi_k|\phi_{k-1}\ra$ of coherent states, one deduces (at first order in $\tau$) the generating functional
\begin{equation}
\label{Z-Doi}
 {\cal Z}[j,\bar{j}] = \displaystyle \int \prod_{\alpha=0}^N 
\frac{d^2 \phi_\alpha}{\pi} \exp\left(-\sum_{k=1}^N \phi_k^*(\phi_k-\phi_{k-1})+\tau \sum_{k=1}^N {\cal W}(\phi_k^*,\phi_{k-1})+j_k\phi_k+\bar{j}_k\phi_k^*\right)
\end{equation}
where the initial conditions and final time contributions have been omitted  since our focus is on stationary states and a loss of memory of initial conditions is assumed (see {\it e.g.} \cite{cardy98} for details). 
Note that the two fields $\phi_k$ and $\phi_k^*$ are  complex conjugate and that we have chosen real sources $j_k,\bar{j}_k $ 
(see Appendix \ref{appendix0}). 

The continuous-time limit consists in taking the limit $\tau\to 0$. In this limit, the discrete sums become integrals over time
and $\phi_k^*(\phi_k-\phi_{k-1})/\tau$ tends to $\phi^*\partial_t\phi$.  As we shall see in the following, this continuous
version of the model becomes ambiguous when the inverse of the operator $\partial_t \delta(t-t')$, that is $\theta(t-t')$,  appears at coinciding times. 
To remove these ambiguities, we  need to refer
to the original time-discretized version of the model, Eq. (\ref{Z-Doi}). 
Let us already mention that the way the coherent states were introduced in the derivation above
leads to the time ordering of  Eq. (\ref{Z-Doi}) where the $\phi^*$ field always appears at a time equal or larger
than the time of the $\phi$ field. We show in the following that this also  occurs for field theories derived from Langevin equations
when  It$\bar{\rm o}$'s time-discretization is chosen. 

An example of a field theory derived from a master equation such as Eq. (\ref{master}) is that
of  branching and annihilating random walks 
in the universality class of directed percolation \cite{cardy98}. A particular model belonging to this class 
 consists in a set of identical particles $A$ diffusing on a lattice
 at rate $D$ and undergoing, upon encounter, the reactions: $A\to 2A$  and  $2A\to \emptyset$ with rates
 $\sigma$ and $\lambda$ and, possibly,  $A\to \emptyset $ with rate $\mu$. 
These models exhibit a phase transition  when the system falls into the  absorbing (empty) state
which sets in, {\it e.g.}, when the branching rate  $\sigma$ is decreased. Obviously, this dynamics 
is irreversible since detailed balance is violated. The transition is thus genuinely out-of-equilibrium.

The reaction-diffusion system above
can be studied using the Doi-Peliti formalism. Its action  writes (in the continuous-space and time limits 
and after a shift and a rescaling of the fields $\phi$ and $\phi^*$):
\begin{equation}
\label{action-dirperco}
{\cal S}[\phi,\phi^*]=\int_{t,\vx} \left(\phi^*\left(\partial_t\phi - D\, \nabla^2\phi\right) +(\mu-\sigma)\phi^*\phi
+\sqrt{2\sigma\lambda}\,\phi^*\phi(\phi-\phi^*)+\lambda(\phi^*\phi)^2\right).
\end{equation}
All field theories derived from this formalism share some common features:
 (i) their actions are proportional  to the field $\phi^*$ (which is an important property, see Section \ref{proportionality}), 
(ii) they are very similar to  the corresponding field theories derived from  Langevin equations (see \cite{benitez11} for a discussion about the correspondence
between the two formalisms).
 We now review the derivation of  field theories ensueing from Langevin equations.

\subsection{Janssen-de Dominicis functional}
\label{janssen}
Suppose that a  field $\phi(t,\vx)$ evolves according to a stochastic dynamics
 described at a mesoscopic level by the Langevin  equation:
\begin{equation}
\partial_t\phi(t,\vx)= - \coefu F(\phi(t,\vx)) +N(\phi(t,\vx))  \zeta(t,\vx)
\label{langevin}
\end{equation}
where $F$ represents the deterministic part of the evolution which can depend on $\phi$
 and its space derivatives, $\coefu$ denotes a constant and uniform relaxation rate, and 
$\zeta(t,\vx)$ is a Gaussian noise:
\begin{equation}
P(\zeta)= \frac{1}{\sqrt{4\pi \Omega_2}}e^{-\frac 1 4 \int_{t,\vx} \zeta^2 /\coefd} \hbox{\hspace{0.5cm}with\hspace{0.5cm}}  \int_{t,\vx}\equiv \int d^d \vx \; dt
\end{equation}
so that 
\begin{equation}
\langle \zeta(t,\vx) \zeta(t',\vx') \rangle = 2 \coefd \, \delta(t-t') \delta^d(x-x')\ .
\end{equation}
 One wants in general to compute averages of functions of the field 
${\cal O}(\phi(t,\vx))$ over the noise distribution:
\begin{equation}
\langle {\cal O}(\phi)\rangle= \int \cD \zeta \,P(\zeta)\, {\cal O}(\phi_\zeta)
\end{equation}
where $\phi_\zeta(t,\vx)$ is the solution of Eq. (\ref{langevin}) for the realization $\zeta$ of the noise. 
This average can be conveniently written as
\begin{equation}
\begin{array}{ll}
\langle {\cal O}(\phi)\rangle &= \int \cD\zeta\, P(\zeta)\,\int \cD\phi\, \delta(\phi-\phi_\zeta)\, {\cal O}(\phi)\\
                       &       \\
                 &= \int  \cD\zeta\, P(\zeta)\,\int \cD\phi\, \delta(\partial_t\phi+ \coefu F(\phi)-N(\phi) \zeta)\, \cal{J}(\phi)\, {\cal O}(\phi)\\
                       &       \\
                 &= \int \cD\zeta\, P(\zeta)\,\int \cD\phi \cD[i\tphi]\, e^{\int_{t,\vx} -\tphi(\partial_t\phi+ \coefu F(\phi)- N(\phi)\zeta) }\,  {\cal J}(\phi)\, {\cal O}(\phi)
\end{array}
\label{average-jacobian}
\end{equation}
 where  the functional analogue of the usual identity
\begin{equation}
\delta(x-x_0)= \delta(f(x))\vert f'(x_0)\vert,
\end{equation}
 with $f(x)$  assumed to have a unique zero $x_0$, has  been used.
In Eq. (\ref{average-jacobian}), the Jacobian $\cal{J}$ writes:
\begin{equation}
{\cal J}(\phi)=\left \vert \rm{det}\left(\partial_t  +\coefu \frac{\delta F(\phi)}{\delta \phi} - \frac{\delta N(\phi)}{\delta \phi} \zeta\right)\right\vert\ .
\label{jacobian}
\end{equation}
Notice that to go from the first to the second equality in Eq. (\ref{average-jacobian}),  
 the initial conditions, which are in principle necessary to solve
the Langevin equation,  have been omitted.  As in the Doi-Peliti formalism, it is implicitly  assumed that the dynamics under study
leads at large time to a steady states with a loss of memory of the initial conditions. This derivation is therefore only 
valid in (or sufficiently close to) the stationary state.

At this stage, two routes can be followed.
The first route consists in introducing a set of (conjugate) Grassmann fields $\eta$ and $\bar\eta$ 
to exponentiate the determinant as 
\begin{equation}
{\cal J}=\int \cD\eta \cD\bar\eta\, e^{\int_{t,\vx} \eta\left( \partial_t +\coefu \frac{\delta F(\phi)}{\delta \phi}  - \frac{\delta N(\phi)}{\delta \phi} \zeta \right)\bar\eta}\ 
\label{jacsusy}
\end{equation}
(assuming that the absolute value plays no role which is the case if there is only one solution
to Eq. (\ref{langevin})). This route is {\it a priori} complicated since it doubles the number of  fields.
It  is in fact convenient only when the system relaxes to equilibrium. In this case, the model satisfies
a supersymmetry that allows  one to reformulate the whole theory in terms of a unique superfield which
makes the  corresponding formalism rather simple. We deal with this case in Section \ref{supersymmetry}.

The second route exploits the identity $\rm{det}= \exp\Tr\ln$, so that the determinant comes 
as an additional term in the exponential in (\ref{average-jacobian}).  This term appears 
to be proportional to the inverse of the operator $(\p_t)\delta(t-t')$, which is 
$\theta(t-t')$, evaluated at $t=t'$ \cite{zinn,honkonen11}. This constant 
$\theta(0) = \epsilon$ is ill-defined because it depends on the precise (discrete) ordering
 of times \cite{janssen92}. The simplest choice $\epsilon=0$ leads to 
${\cal J}=1$ and corresponds in discrete time to It$\bar{\rm o}$'s discretization choice.
Once the  Jacobian is set to unity, one can finally integrate in (\ref{average-jacobian}) 
over the Gaussian noise distribution and deduce
the generating functional for correlation and response functions
\begin{equation}
{\cal Z}[j,\tilde{j}]=\int \cD\phi\, \cD [i\tphi]\, e^{-{\cal S}[\phi,\tphi]+\int_{t,\vx}j\phi+\tilde{j}\tphi}
\label{zlangevin}
\end{equation}
with
\begin{equation}
{\cal S}[\phi,\tphi] =  \int_{t,\vx} \tphi\left(\partial_t\phi+ \coefu F(\phi)\right) - \coefd N^2(\phi)\tphi^2 .
\label{action-lang}
\end{equation}

As in the case of reaction-diffusion systems, and for the same reasons, the ambiguity on the value
of $\theta(0)$ has to be removed both perturbatively and nonperturbatively to make consistent
calculations from Eqs. (\ref{zlangevin},\ref{action-lang}). Since this ambiguity is related to
the continuous-time limit we first review the discrete time
version of the action (\ref{action-lang}).

\subsection{ Field theory from discrete-time  Langevin equations}
\label{discLangevin}

 The Langevin equation (\ref{langevin}) in  It$\bar{\rm o}$'s  discretization reads (assuming for simplicity $\Omega_1=\Omega_2=1$)
\begin{equation}
 \phi_{n}-\phi_{n-1} = \tau F(\phi_{n-1}) + N(\phi_{n-1})\zeta_{n-1}
\end{equation}
where $n$, $n-1$ are time indices and where $\tau$ denotes the time step. Any reference to the space dependence
has been omitted here since it plays no role in the following discussion.
With this stochastic process is associated a transition probability
\begin{equation}
 T_\tau(\phi_{n}|\phi_{n-1}) = \displaystyle \left(2\pi N^2(\phi_{n-1})\tau\right)^{-1/2}\exp\left(-\frac{\left(\phi_{n}-\phi_{n-1} - 
\tau F(\phi_{n-1})\right)^2} {2 N^2(\phi_{n-1})\tau}\right)
\end{equation}
that can be rewritten as the following Gaussian integral
\begin{equation}
  T_\tau(\phi_{n}|\phi_{n-1}) = \displaystyle \int_{-i\infty}^{i\infty} \frac{d \tphi_{n}}{2\pi i} \Omega_\tau(\phi_{n},\tphi_{n}|\phi_{n-1})
\end{equation}
with
\begin{equation}
 \Omega_\tau(\phi_{n},\tphi_{n}|\phi_{n-1}) =\exp\left(-\tphi_{n}( \phi_{n}-\phi_{n-1} - \tau F(\phi_{n-1})) + \tau N^2(\phi_{n-1})\tphi_{n}^2 \right)
\end{equation}
where $\tphi_{n}$ is the conjugate, or response, variable associated with the transition from $\phi_{n-1}$ to $\phi_{n}$.
For a Markov chain of $N$ transitions between times 0 and $t_N=\tau N$, 
the total transition probability is the product
\begin{equation}
 P_N(\phi_1,t_1;\dots;\phi_N,t_N|\phi_0,0) = \displaystyle \prod_{n=1}^{N}T_\tau(\phi_n|\phi_{n-1}).
\end{equation}
Hence, one obtains the generating functional of correlation and response functions \cite{janssen92}: 
\begin{equation}
\begin{array}{l c l}
 Z[\{j_n,\tJ_n\}] &=& \displaystyle \frac{1}{2\pi i}\int \prod_{n=1}^N d \phi_n  d\tphi_n e^{-{\cal S}[\phi,\tilde\phi]+ \sum_{n=1}^N j_n \phi_n +\tJ_n\tphi_n}\\
\hbox{with} & & \displaystyle{\cal S}[\phi,\tilde\phi] = \tau \sum_{n=1}^N \tphi_n\left((\phi_n-\phi_{n-1})/\tau -F(\phi_{n-1})\right) -N^2(\phi_{n-1})\tphi_n^2.
\label{actdis}
\end{array}
\end{equation}
We emphasize that  in this action all the $\tphi$ fields  appear at a time larger or equal  than the times of all the $\phi$ fields
as in the Doi-Peliti procedure for reaction-diffusion systems, Eq. (\ref{Z-Doi}). The ambiguity in the value
of $\theta(0)$ is therefore the same in both cases.
Before embarking into the NPRG  formalism let us first recall (and reformulate in a way useful for NPRG) 
how It$\bar{\rm o}$'s prescription is dealt with in perturbation theory.

\subsection{Dealing with  It$\bar{\rm o}$'s discretization in perturbation theory}
\label{perturbative-ito}
To identify the relevant prescriptions, we use the time-discretized version of the field theory
 (\ref{actdis}), which is the only form free from ambiguity. 
As mentioned above, the time ordering and the structure of the Janssen-de Dominicis
 and of the Doi-Peliti functionals are the same. Hence the considerations
 developed below hold for both functionals.

We consider the quadratic part of the action
\begin{equation}
 {\cal S}_0[\phi,\tilde\phi] = \tau \sum_{n=1}^N \tphi_n \left(\phi_n-\phi_{n-1}\right)/\tau + w \tphi_n \phi_{n-1},
\end{equation}
with typically $w=\nabla^2 + m^2$, and denote ${\cal Z}_0$ the corresponding generating functional. 
Integrating over $\tilde\phi$ in ${\cal Z}_0$ produces a product of delta functions which enforces the relation
\begin{equation}
 M.{}^t(\phi_0 \dots \phi_N) =  {}^t(\tJ_0 \dots \tJ_N)
\end{equation}
where $^t(\dots)$ is the transpose operator, and $M$ is the $N\times N$ matrix with components 
$M_{i j } = \delta_{i,j} + w \delta_{i-1,j}$. This matrix $M$ is precisely the matrix whose  determinant  
appears in Eq. (\ref{jacobian}). It is clear from its definition that det$(M)=1$ and, since  
 this result straightforwardly generalizes to any interacting theory, this proves
that in  It$\bar{\rm o}$'s discretization  the Jacobian is unity.
Inverting this matrix: $M^{-1}_{ij} = \delta_{i j} + \sum_{k=1}^{N-1} (-w)^{k} \delta_{i-k, j}$
 allows one to perform the remaining integrations over the $\phi_n$ and one finds
\begin{equation}
 {\cal Z}_0[\{j_n,\tJ_n\}] = \exp\left({\displaystyle\sum_{m,n=1}^N j_m\; M^{-1}_{m n}\;\tJ_n}\right).
\end{equation}
One can then compute the free two-point response function (the bare propagator $G_0$)
\begin{equation}
\label{free-propag}
 G_{0 ,kl}= \la  \phi_k \tphi_l \ra =\frac{\delta^2 \ln Z_0}{\delta j_k \delta \tJ_l} = M^{-1}_{k l} = \theta(l \le k) (-w)^{k-l}
\end{equation}
which vanishes if the  field $\phi$ appears at an earlier time than the field $\tphi$ as a signature of causality.
It is important  to notice that $\la  \phi_k \tphi_k \ra=1$ whereas $\la  \phi_k \tphi_{k+1} \ra=0$ 
which is the reason why the continuous-time limit $\tau\to 0$ is delicate.

\begin{figure}
\begin{center}
\includegraphics[width=42mm,height=26mm]{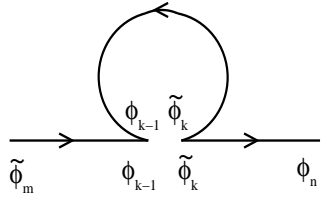}
\end{center}
\label{fig1}
\caption{ Example of a perturbative tadpole diagram. The time grows from left to right and the arrows go from $\tilde{\phi}$ to $\phi$, allowing to keep track of causality. In this example, the interaction term in the action is  $ \tphi^2 \phi^2$ and the time separation  between the $ \tphi$ and the $\phi$ fields at the vertex is the time step $\tau$ of It$\bar{\rm o}$'s discretization. The graph is vanishing because the arrow in the tadpole is going backward in time. }
\end{figure}

In perturbation theory, the ambiguity occurs only in tadpoles in which the bare propagator $G_0(t-t')\propto\theta(t-t')$ starts 
and ends at the same vertex. The value of a tadpole is therefore proportional to $G(0)\propto\theta(0)$ which is  ill-defined. 
Bearing in mind 
that in discrete time the interaction vertices come with the fields shifted in time
according to  Eq. (\ref{actdis}), that is, are of the form 
 $\tphi^n_k \phi_{k-1}^m$, one concludes that the bare vertices actually involve two times 
separated by the time step $\tau$. As a consequence, the propagator appearing in a 
tadpole -- joining a $\phi$ with a $\tphi$ -- is thus $\la  \phi_{k-1} \tphi_{k} \ra$,  which is 
vanishing because of (\ref{free-propag}), and not $\la  \phi_k \tphi_k \ra$   (see Fig. 1). This result,
  grounded on the analysis of the discrete time  field theories, implies that in the continuum the propagator  in 
 a tadpole is $\la  \phi(t)\tphi(t+\tau) \ra_0=G_0(t-(t+\tau))=G_0(-\tau)$
and that therefore $\theta(0)$ should be understood as:
\begin{equation}
 \theta(0) = \lim_{\tau\to 0}\theta(-\tau) = 0.
\label{theta-ito}
\end{equation}   
This yields the standard prescription used in perturbation theory: in It$\bar{\rm o}$'s discretization, 
the $\theta(0)$ appearing in $G_0(0)$ must be set to 0 in tadpoles.

Let us notice that the need for a prescription  in perturbation theory is directly related to the fact that in continuous time
the bare vertices are local in time and thus are singular functions of their time arguments (Dirac functions such as $\delta(t-t')$).
 The interesting  point for what follows is that, since only {\it the product} of a bare propagator and of a bare vertex matters,
 one can remove the ambiguity, either by considering 
that in the vertices the $ \tilde\phi$ fields are shifted in time with respect to the $\phi$ fields, 
 or by considering that this shift occurs in the propagator. In this second solution, 
It$\bar{\rm o}$'s prescription Eq. (\ref{theta-ito}) is effectively implemented
 by shifting the time $t'$ of $\tilde\phi(t')$ in $ G_0(t-t')$, that is,
by replacing  $\la  \tilde\phi(t',\vx')\phi(t,\vx)\ra_0$ by
\begin{equation}
 \la  \tilde\phi(t',\vx')\phi(t,\vx)\ra_{0,\epsilon} \equiv \la  \tilde\phi(t'+\epsilon,\vx')\phi(t,\vx)\ra_0
\end{equation}
with $\epsilon\to 0^+$ or, equivalently, in Fourier space
\begin{equation}
 \la  \tilde\phi(\omega', \vq')\phi(\omega, \vq)\ra_{0,\epsilon}\equiv e^{-i\epsilon \omega'} \la  \tilde\phi(\omega', \vq')\phi(\omega, \vq)\ra_0.
\label{propag-shift}
\end{equation}
Doing so precisely amounts to changing in tadpoles $G_0(0)$  by $G_0(-\epsilon)$, that is, 
$\theta(0)$ by  $\theta(-\epsilon)$ as it should according to Eq. (\ref{theta-ito}).

\section{The non-perturbative renormalization group for out-of-equilibrium models }
\label{nprg}

The general idea underlying the NPRG  is the same  for equilibrium and  
out-of-equilibrium problems: one
builds a sequence of effective models that interpolate smoothly between the micro- 
and the macro-physics
and that consist, at an intermediate length-scale, in the integration up to this scale 
over the stochastic fluctuations.
Two important features of out-of-equilibrium problems must nevertheless be specifically
considered, the doubling of each field by a response field, and causality. 

\subsection{General NPRG formalism for out-of-equilibrium models}

We  consider the field theory given by Eq. (\ref{actdis}) in It$\bar{\rm o}$'s discretization (or  Eq. (\ref{Z-Doi})
for reaction-diffusion systems).
As we shall show in the following, this time-discretization requires to implement consistently
a prescription on the propagator of the theory. We ignore this for the moment as we will show
in  Section \ref{ito} that it can be easily taken into account  {\it a posteriori}.

As in equilibrium, one wants to generate, through progressively averaging over fluctuations, 
a sequence of  scale-dependent models, whose effective actions  $\Gamma_{\kappa}$
   interpolate smoothly between the bare action  $ {\cal S}$ of the initial model 
and its effective action $\Gamma$, that is,
 the (\nequ analogue of the) Gibbs free energy. This requires that at the microscopic scale  $\Lambda$
(the inverse of the lattice spacing for instance) $\Gamma_{\kappa=\Lambda }= {\cal S}$
and that at scale $\kappa=0$ , $\Gamma_{\kappa=0}=\Gamma$.
At an intermediate scale $\Lambda>\kappa>0$, the slow modes (with respect to $\kappa$) are  decoupled {\it by construction}
 in order to make $\Gamma_\kappa$  (almost) equal to the effective action of the rapid modes.

The decoupling of the slow modes is achieved by adding  a quadratic (mass-like) momentum-dependent  term
$\Delta{\cal S}_\kappa$ to the original action.
The required properties of this term are (i) to be very large at scale  $\kappa=\Lambda$ for all momentum modes
 (so that all  stochastic fluctuations are frozen and the mean field becomes exact:
 $\Gamma_\Lambda = {\cal S}$  -- see Appendix \ref{appendix0}), (ii) to vanish at scale  $\kappa=0$ (so that the original model 
is recovered: $\Gamma_{\kappa=0}=\Gamma$) and  (iii) to give 
a ``square mass'' of order $\kappa^2$ to the slow modes
when $\Lambda>\kappa>0$ in order to decouple them from the long distance physics (integration over the rapid modes only). 
 We therefore build, as in equilibrium,
 a scale-dependent generating functional of the correlation and response functions:
\begin{equation}
\label{part-function}
Z_\kappa[j,\tilde{j}]=\int \cD\phi \cD\tilde\phi 
\exp\left(-{\cal S}-\Delta{\cal S}_\kappa+\int_{\bf x}  { }^tJ({\bf x}).\Phi({\bf x})\right)
\end{equation}
with  ${\bf x}=(t,\vec{x})$ 
\begin{equation}
\Phi({\bf x})=\left(
\begin{array}{l}
\phi({\bf x})\\
\tilde\phi({\bf x})
\end{array}
\right)\ \ {\rm and}\ \ 
J({\bf x})=\left(
\begin{array}{l}
j({\bf x})\\
\tilde{j}({\bf x})
\end{array}
\right)
\end{equation}
 and
\begin{equation}
\Delta{\cal S}_\kappa=\frac{1}{2}\int_{{\bf x},{\bf x'}} { }^t\Phi({\bf x}). \hat{R}_\kappa({\bf x}-{\bf x'}).\Phi({\bf x'})
\end{equation}
where $\hat{R}_\kappa$ is the $2\times2$ matrix of  mass-like cutoff functions that suppress the fluctuations of the slow modes
(see Appendix \ref{appendixA} for the definitions used throughout the text).

We also define as in equilibrium the generating functional of connected functions 
$W_\kappa [J]=\log Z_\kappa[J]$ and its (modified) Legendre transform $\Gamma_\kappa$  by
\begin{equation}
\Gamma_\kappa[\Psi]+W_\kappa [J]=\int_{\bf x}   { }^tJ.\Phi -
\frac{1}{2}\int_{{\bf x},{\bf x'}} { }^t\Psi({\bf x}). \hat{R}_\kappa({\bf x}-{\bf x'}).\Psi({\bf x'})
\end{equation}
where
\begin{equation}
\Psi=\la \Phi\ra.
\end{equation}
It is convenient to define two  notations for vertex (one-particle-irreducible) correlation functions:
\begin{equation}
\Gamma^{(n)}_{\kappa;i_1,\dots,i_n} [\{{\bf x}_i \};\Psi]=
\frac{\delta^n \Gamma_\kappa [\Psi]}{\delta \Psi_{i_1}({\bf x}_1)\dots\delta \Psi_{i_n}({\bf x}_n)}
\end{equation}
 and
\begin{equation}
\Gamma^{(n,\tilde{n})}_{\kappa} [\{{\bf x}_i \},\{{\bf x'}_j \};\Psi]=
\frac{\delta^{n+\tilde{n}}\Gamma _\kappa [\Psi]}{\delta \psi({\bf x}_1)\dots\delta\tilde{\psi}({\bf x'}_{\tilde{n}})}
\end{equation}
where $\{{\bf x}_i \}$ stands for $({\bf x}_1,\dots{\bf x}_n)$ and  $\{{\bf x'}_j\}$ for  
$({\bf x'}_1 ,\dots, {\bf x'}_{\tilde{n}} )$ (see Appendix \ref{appendixA} for the  
precise correspondence between both notations).

The exact flow for $\Gamma_\kappa$ is given by Wetterich's equation \cite{tetradis94,berges02}
\begin{equation}
\partial_\kappa \Gamma_\kappa[\Psi]  = 
\displaystyle \frac{1}{2}\, \Tr\int_{{\bf x},{\bf x'}}\! \partial_\kappa \hat{R}_\kappa ({{\bf x}-{\bf x'}}).\hat{G}_\kappa[{{\bf x},{\bf x'}};\Psi]
\label{dkgam}
\end{equation}
with $\hat{G}_\kappa \equiv\left[\hat{\Gamma}_\kappa^{(2)}+\hat{R}_\kappa\right]^{-1}$  the 
full field-dependent  propagator and where
$\hat{\Gamma}_\kappa^{(2)}$  is the $2\times2$ matrix whose  elements  are the  $\Gamma^{(2)}_{\kappa;ij}$.

The flow equation of the two-point functions evaluated in a uniform (in time and space) field 
configuration $\Psi_u$ is also particularly important. It is trivially derived from Eq. (\ref{dkgam}) and can be written as,  in Fourier space 
\begin{equation}
\begin{array}{l}
\displaystyle {\partial_\kappa \Gamma^{(2)}_{\kappa,ij} ({\bf p};\Psi_u)
 = {\rm Tr} \int_{\bf q} \partial_\kappa\hat{R}({\bf q}).\hat{G}({\bf q}) . }\\
\displaystyle {\left( 
\hat{\Gamma}_i^{(3)}({\bf p},{\bf q}).\hat{G}({\bf p+q}).\hat{\Gamma}_j^{(3)}(-{\bf p},{\bf p+q})
 -  \frac{1}{2}\,\hat{\Gamma}_{ij}^{(4)}({\bf p},{-\bf p},{\bf q})\right).\hat{G}({\bf q})}
\end{array}
\label{dkgam2ito}
\end{equation}
where ${\bf p}=(\nu,\vec{p})$ represents a couple (frequency, momentum). 
The $\Psi_u$ dependence is implicit in the right hand side of Eq. (\ref{dkgam2ito}) 
to alleviate the notations and $\hat{\Gamma}_i^{(3)}$ 
(resp. $\hat{\Gamma}_{ij}^{(4)}$) is the $2\times2$ matrix of functional derivatives of 
$\hat{\Gamma}^{(2)}$ w.r.t. $\Psi_i$ (resp.  w.r.t. $\Psi_i$ and $\Psi_j$)  evaluated in
the configuration  $\Psi_u$, (see Appendix \ref{appendixA} for more details and notations).

Of course, as in equilibrium, such a flow equation is not closed, and is only the
starting point of an infinite hierarchy. Approximations are needed to obtain any useful result.
The most widely used so far is the  derivative expansion, 
where the vertex functions are, in Fourier space, 
expanded as a power series of their momenta and frequencies and which we will use
in our application of NPRG to Model A in Section \ref{modelA}.

\section{It$\bar{\rm o}$'s prescription and NPRG}
\label{ito}

The flow of a vertex function
 involves integrals of products of other vertex functions $\Gamma_\kappa^{(n,\tilde{n})}$  and of full 
propagators $G_\kappa$ (see {\it e.g.} Eq. (\ref{dkgam2ito})).
  As in perturbation theory, an ambiguity can only appear
when two times coincide. If the (full) vertex functions $\Gamma_\kappa^{(n,\tilde{n})}$  are
smooth functions of their time arguments, no ambiguity can arise  because the 
times $t$ and $t'$ of the two ends of a propagator $G_\kappa(t-t')$ joining two legs of $\Gamma_\kappa^{(n,\tilde{n})}$
  is integrated over so that the value of $G_\kappa(0)$  becomes immaterial in these integrals. 
Once again, this value plays a role only if the propagator  $G_\kappa(t-t')$ is multiplied by a singular
function such as, for instance, a Dirac function $\delta(t-t')$.

 This is precisely what happens within 
the derivative expansion  where the vertex functions are, in direct space, expanded as 
(formal) power series  of Dirac functions and of their derivatives: $\delta^{(m)}(t-t')$. 
As in the perturbative case and for the same reason, 
 the trick to get rid of the ambiguities is
to shift in the full propagator $G_\kappa$ (actually in $G_\kappa^{(1,1)}=\la \phi \tilde\phi \ra$ which is the only causal two-point function)
 the time of the $\tilde\phi$ field by an infinitesimal amount 
$\epsilon$. Before proceeding, let us prove a theorem.

\subsection{Causality of the response functions and choice of cutoff functions}
The causality property of Eq. (\ref{free-propag}) can be straightforwardly generalized 
 to all  response functions both for the free field theory and for interacting 
theories truncated at the mean field (tree) level \cite{janssen92}. In 
discrete time it writes:
\begin{equation}
 \left\la \prod_{i=1}^n \phi_{k_i} \prod_{j=1}^{\tilde{n}} \tilde\phi_{l_j}\right\ra =0 \hbox{\;\;\;\;\; 
if\  \ $\exists \; l_j$/ $\forall k_i$, $l_j>k_i$ }
\label{causal}
\end{equation}
for $n>0$ and $ \tilde{n}>0$. This equality means that for a response function   to be nonzero 
the largest time must be that of a $\phi$ field.
We now prove (in continuous time) that this equality holds at any scale 
$\kappa$ of the NPRG flow.  In fact, to preserve this property requires to appropriately choose  the cutoff functions
$\hat{R}_\kappa$ and we now discuss this point.

 Most often, it  turns out to be sufficient for the decoupling of the slow modes to choose an anti-diagonal matrix  $\hat{R}_\kappa$,
that is, a $\Delta{\cal S}_\kappa$  term that couples only $\phi$ with 
$\tilde\phi$. The exception to this rule  arises when the symmetries of the model, that should of course be preserved by the cutoff
term,  enforce the presence of an additional (diagonal) term,  proportional to $\tilde\phi^2$ (as for instance  
 occurs in the field theory associated with  the Kardar-Parisi-Zhang equation \cite{canetcond,canet10}). To the best of our knowledge,
a cutoff term proportional to $\phi^2$ is never necessary and we do not consider it in the following (it  would become problematic
for the proof of the property (\ref{causal}) and of the fact that $\Gamma_\kappa$ is proportional to $\tilde\psi$, 
see Section \ref{proportionality}). 

The cutoff term coupling   $\phi$ with  $\tilde\phi$ must of course also preserve causality.
It is therefore either a function ``independent'' of time (actually of frequency): 
$R_\kappa(t-t',\vx-\vx')\to\delta(t-t'){R}_\kappa(\vx-\vx')$ or  proportional to $\theta(t-t')$.
For the sake of simplicity we consider in the following  a ``time-independent'' cutoff
function  but the generalization to a general causal cutoff
would be  straightforward.  To further simplify, we consider only a  $\Delta{\cal S}_\kappa$  term
that couples $\phi$ with $ \tilde\phi$ although the following arguments can be  
rather straightforwardly   generalized to an additional cutoff term proportional to $ \tilde\phi^2$.  
 We finally choose a cutoff term with a $\tilde\phi$ shifted in time:
\begin{equation}
\Delta{\cal S}_\kappa=\int_{t,\vx,\vx'} R_\kappa(\vx-\vx')\;\phi(t,\vx)\tilde\phi(t+\epsilon,\vx\,').
\end{equation}
 
The proof  that (\ref{causal}) is preserved at every scale $\kappa$ is made by induction. First, the property is obviously satisfied
at scale $\kappa=\Lambda$ since the mean field is the initial condition of the NPRG flow:
$\Gamma_\Lambda= \cal{S}$ (for a proof of this equality, see Appendix \ref{appendix0}).  
Let us suppose that it holds at a scale $\kappa$.
Then, at scale $\kappa-d\kappa$ the property is preserved if the variation of the 
Green functions coming from the RG flow also satisfies  Eq. (\ref{causal}). The starting point 
of the proof is the NPRG  flow equation for ${\cal W}_\kappa$ that writes:
\begin{equation}
\partial_\kappa W_\kappa=\int_{t,\vx,\vx'}\partial_\kappa R_\kappa(\vx-\vx')\;\la \phi(t,\vx)\tilde\phi(t+\epsilon,\vx')\ra_\kappa
\label{flot-NPRG-Z}
\end{equation}
 where the index $\kappa$ in $\la \dots\ra_\kappa$ means 
that the average is taken in presence of the $\Delta{\cal S}_\kappa$ term. The flows of the 
connected functions follow  from Eq. (\ref{flot-NPRG-Z}):
\begin{equation}
\partial_\kappa \frac{\delta^{n+\tilde{n}} W_\kappa}{\delta j_{1}\dots\delta\tilde{j}_{\tilde{n}}}=
\int_{t,\vx,\vx'}\partial_\kappa R_\kappa(\vx-\vx')\;\la \phi_1\dots\tilde\phi_{\tilde{n}}\;\phi(t,\vx)\tilde\phi(t+\epsilon,\vx')\ra_\kappa
\label{flot-NPRG-connected}
\end{equation}
where the indices $i$ stand for $(t_i,\vx_i)$. By hypothesis, the function
appearing on the right  hand side of Eq. (\ref{flot-NPRG-connected})  is
nonvanishing only when its largest time  is that of a $\phi$ field. It
cannot be  $\phi(t,\vx)$  since $\tilde\phi(t+\epsilon,\vx')$
is posterior  and it is therefore  one of the other
$\phi_i$   fields.     This   proves   that    the   contribution   to
$W_{\kappa}^{(n,\tilde{n})}$ of the momentum shell $d\kappa$
is nonvanishing  only if its  largest time is that  of a
$\phi$ field. By iteration from  the inital condition we conclude that
the property Eq. (\ref{causal}) holds  for any $\kappa$. 

Note that a similar result can be  derived  for the one-particle-irreducible 
vertex functions $\Gamma_{\kappa}^{(n,\tilde{n})}$ with the difference that the latest
time must be that of a $\tilde\psi$ field (this subtlety  coming from the fact
that these functions are amputated of the propagators of their external legs).

\subsection{It$\bar{\rm o}$'s prescription on the full propagator $G_{\kappa}^{(1,1)}(t,t')$ and on  $\Gamma_\kappa$ }
\label{proportionality}
The  previous result implies in
particular    that   the    running   full    (connected)   propagator
$G_{\kappa}^{(1,1)}(t,t')=\la  \tilde\phi(t',\vx')\phi(t,\vx)\ra=W_{\kappa}^{(1,1)}(t,t')$    remains   proportional   to
$\theta(t-t')$ all  along the  flow since it is nonvanishing only when $t$ is larger than $t'$.  The ambiguity  at coinciding times
  remains   therefore  identical  to   the  one
encountered perturbatively.  As emphasized in  Section \ref{perturbative-ito}, 
  the ambiguity can be equivalently removed in perturbation theory by shifting  the $\tilde\phi$
fields at  the vertices or in  the bare propagator.  This second way of  
shifting time can in fact be formulated at the level of the 
discretized theory and not only
in  the  perturbative  expansion, therefore it can be exploited also nonperturbatively.   The  way  to  keep  track  of  a shift
  of time  in the  propagator $G_{\kappa}^{(1,1)}(t,t')$ while
Fourier transforming  is to modify  it as was done in
the perturbative case on   $G_{0}(t,t')$,  Eq. (\ref{propag-shift}). The nonperturbative 
It$\bar{\rm o}$'s prescription is therefore to replace the full propagator  $G_{\kappa}^{(1,1)}(t,t')$ by
\begin{equation}
 \la  \tilde\phi(t',\vx')\phi(t,\vx)\ra_{\epsilon} \equiv \la  \tilde\phi(t'+\epsilon,\vx')\phi(t,\vx)\ra
\end{equation}
with $\epsilon\to 0^+$ , that is to multiply it in Fourier space by the regularization factor exp$(-i\epsilon\omega')$:
\begin{equation}
 \la  \tilde\phi(\omega', \vq')\phi(\omega, \vq)\ra_{\epsilon}
\equiv e^{-i\epsilon \omega'} \la  \tilde\phi(\omega', \vq')\phi(\omega, \vq)\ra.
\label{propag-shift-nonpert}
\end{equation}
Note that when this function is evaluated in a uniform field configuration, it becomes
 proportional to $\delta(\omega+\omega')$ in which case $\exp (-i\epsilon \omega')=\exp ( i\epsilon \omega )$.\\

Let us finally prove the following theorem which is a consequence of It$\bar{\rm o}$'s prescription:
if the bare action of a \nequ system and the regulator term $\Delta {\cal S}_\kappa$ are proportional to $\tilde\psi$, then the effective average action $\Gamma_\kappa[\psi,\tilde\psi]$ is also proportional to $\tilde\psi$. The proof is as follows. 
Let us rewrite $\Gamma_\kappa[\psi,\tilde\psi]$ as the sum of a term independent of
$\tilde\psi$ and of a term proportional to $\tilde\psi$:
\begin{equation}
\label{def-gamma-V}
 \Gamma_\kappa[\psi,\tilde\psi]=\int_{\bf  x}\left(  V_\kappa(\psi({\bf  x}) ) + \tilde\psi({\bf  x}) \Gamma_\kappa^{(0,1)}[{\bf  x};\psi,\tp]\right).
\end{equation}
Then:
\begin{equation}
\label{def-V}
 \int_{\bf  x} V_\kappa(\psi({\bf  x}))= \Gamma_\kappa[\psi,0].
\end{equation}
By hypothesis,  $V_{\kappa=\Lambda}=0$ since  $\Gamma_{\kappa=\Lambda}=  {\cal S}$ (see Appendix \ref{appendix0}) and $ {\cal S}$
is proportional to $\tilde\psi$. Let us now suppose that  $V_{\kappa}$ remains zero down to the scale 
$\kappa_0$. Then at $\kappa_0 - d\kappa$, $V_{\kappa_0 - d\kappa}$ is vanishing if and only if the contribution 
to $V_{\kappa_0 - d\kappa}$  coming from the flow between $\kappa_0$ and $\kappa_0 - d\kappa$ is zero. 
This contribution is calculated from
\begin{equation}
\label{flot-V-psi}
\begin{array}{ll}
 \partial_\kappa\displaystyle{\int_{\bf  x}  V_\kappa(\psi({\bf  x}))_{\vert_{\kappa_0}}}& \displaystyle{=  \partial_\kappa\Gamma_\kappa[\psi,0]{\Big\vert_{\kappa_0}}}\\
                &      \displaystyle{=\frac{1}{2}\Tr\int_{\bf q}\partial_\kappa \hat{R}_\kappa({\bf q}) \hat{G}_\kappa[{\bf q},-{\bf q};\psi,\tilde\psi=0]{\Big\vert_{\kappa_0}}}.
\end{array}
\end{equation}
{In this equation, $\hat{G}_{\kappa}$
 is  computed by inverting the matrix $(\hat\Gamma^{(2)}_{\kappa_0} +\hat R_{\kappa_0})$. As  $V_{\kappa_0}(\psi)=0$,
 the element $\Gamma_{\kappa_0} ^{(2,0)}[{\bf q},-{\bf q}]$ of  $\hat\Gamma^{(2)}_{\kappa_0}$ is vanishing 
  at $\tilde\psi=0$. Moreover, recalling that, by assumption, $\Delta {\cal S}_\kappa$ does not have any $\phi\phi$ term (that is,  as in definition (\ref{regul_app}) of Appendix \ref{appendixA}, $R_\kappa^{20}=0=\p_\kappa R_\kappa^{20}$), 
  the matrix element of the second row and second column  of $\hat{G}_\kappa$  is  vanishing. 
Therefore, one obtains:} 
\begin{equation}
\label{flot-V-ito}
\begin{array}{ll}
\displaystyle{\int_{\bf  x} \partial_\kappa V_\kappa(\psi({\bf  x})){\Big\vert_{\kappa_0}}}=&\displaystyle{\frac{1}{2}
\int_{\bf q}\partial_\kappa \hat{R}^{11}_{\kappa_0} ({\bf q}) }\\
&\displaystyle{\left(  
{e^{-i\epsilon\omega}}{G_{\kappa_0} ^{(1,1)}[{-\bf q},{\bf q}]} + {e^{i\epsilon\omega}}{G_{\kappa_0} ^{(1,1)}[{\bf q},-{\bf q} ]}\right)}.
\end{array}
\end{equation}
Now, because of the causality of $G_{\kappa}^{(1,1)}[{-\bf q},{\bf q}]$, its poles  
lie in the upper half complex plane of $\omega$. The regulator term $\exp(-i\epsilon\omega)$ in 
the integral of the first term allows for  the closure of the integration
contour by a semi-circle at infinity in the lower half plane without changing the value of  the integral. From the residue theorem, we conclude
that the integral over $\omega$ is vanishing since no pole is enclosed in the integration contour.
  The same holds true for the  integral
of the second term
and we therefore conclude that $V_{\kappa_0 - d\kappa}$ is vanishing. By iteration,  at all scales, $V_{\kappa }=0$. 

Note that, had we neglected the regulator terms  $\exp(\pm i\epsilon\omega)$ in the integrals above, 
we would not have found a vanishing flow for $V_{\kappa}$ and we would have concluded incorrectly 
that this term was generated by the RG flow. Note also that once $\Gamma_\kappa$
 has been rewritten as in Eq. (\ref{def-gamma-V}) with $V_\kappa=0$, its flow is entirely 
determined by that of $\Gamma_\kappa^{(0,1)}$ which is better
 behaved because it involves two propagators -- as can be checked by taking the derivative 
of  Eq. (\ref{dkgam}) w.r.t. $\tilde\psi$. It turns out that in all the systems
 studied so far ({\it e.g.} \cite{canet04,canet10}), the resulting integrals 
were  unambiguous and the regulator terms $\exp(\pm i\epsilon\omega)$  not necessary. 
We thus conjecture that the shortcut for It$\bar{\rm o}$'s prescription
 is to impose that $\Gamma_\kappa$ is proportional to $\tilde\psi$ 
(at least if $\Gamma_\kappa$ is an analytic functional of $\tilde\psi$).
Let us now show how the NPRG formalism can be used to study   Model A.

\section{The derivative expansion at order two applied to Model A}
\label{modelA}

Model A describes the purely dissipative relaxation of a non-conserved scalar field $\phi(t,\vx)$ with Ising symmetry. 
It corresponds to Glauber dynamics (single spin flips).
 The model is defined by the Langevin equation (\ref{langevin}) with $N(\phi)=1$ and  $F$ deriving from the standard 
(equilibrium) $\phi^4$   Hamiltonian
\begin{equation}
\label{H-ising}
H[\phi]=\int_{t,\vx} \frac{1}{2}(\nabla\phi)^2 + V(\phi) \hbox{\hspace{0.5cm}with\hspace{0.5cm}}
V(\phi) = \frac{r}{2} \phi^2 + \frac{u}{4!}  \phi^4,
\end{equation}
with the usual $Z_2$ symmetry.
 The action of the model hence writes
\begin{equation}
{\cal S}[\phi,\tphi] =  \int_{t,\vx} \left\{\tphi\left(\partial_t\phi -  \nabla^2\phi+ V'(\phi)\right) -  \tphi^2 \right\} .
\label{actA}
\end{equation}
When approaching the continuous phase transition, the relaxation time of the order parameter
  starts diverging, which reflects  the critical slowing down of the dynamics.
Besides the static critical exponents $\nu$ and $\eta$ of the Ising universality class,
  the critical dynamics  is characterized  by the dynamical exponent $z$
  which relates the divergences of the
  relaxation time $\tau$ and of the correlation length $\xi$ in the vicinity 
of  the critical point as $\tau\sim \xi^z\sim |T-T_c|^{-z\nu}$
 where $T_c$ is the critical temperature.

 In the long-time limit this relaxation-towards-equilibrium model shows a time reversal symmetry that can be  expressed 
 as an invariance of the action (\ref{actA}) under the following field transformation \cite{biroli05,canet07}
\begin{equation}
\left\{
\begin{array}{l l l}
t & \to& -t\\
\phi & \to & \phi \\
\tphi & \to & \tphi - \p_t\phi.
\end{array}
\right.
\label{trs}
\end{equation}
Indeed,  on the one hand, the  equilibrium part $F = {\delta {\cal H}}/{\delta \phi}$ of the action (\ref{actA}) is invariant on its own under the transformation (\ref{trs}) since the additional term $\propto \p_t \phi \,{\delta {\cal H}}/{\delta \phi}$   vanishes upon time integration  in the stationary regime. 
On the other hand, the time-evolution  $\tphi \p_t\phi$ and the noise $\tphi^2$ parts are not invariant on their own  but the combination $\tphi \p_t \phi - \tphi^2$ is,  since the terms 
 generated by the transformation (\ref{trs}), that are  proportional to  
$(\p_t \phi)^2$ and $ \tphi\p_t \phi$,  cancel out or combine to give back the original terms of the action.
 We refer the reader to Ref. \cite{biroli05} for a detailed and general  study  of the field-theoretic formulation of the time reversal symmetry.

The computation of the static critical exponents of  the Ising model has been performed in all 
dimensions using the derivative expansion \cite{berges02,canet04}. 
We now show how to compute the dynamical exponent $z$  within this framework (using a richer  approximation than in the former 
calculation of Ref. \cite{canet07}).

The exact flow equations
of the correlation and response functions do not form a closed set of equations since the flow of $\Gamma^{(2)}_{\kappa}$ 
involves  $\Gamma^{(3)}_{\kappa}$
and  $\Gamma^{(4)}_{\kappa}$ whose  flow equations involve in turn $\Gamma^{(5)}_{\kappa}$
and  $\Gamma^{(6)}_{\kappa}$, {\it etc}. As already mentioned, 
the derivative expansion is probably the simplest and most popular 
approximation -- in particular for the study of  critical properties -- 
to close this hierarchy of equations. 
It amounts to proposing an {\it Ansatz} for $ \Gamma_{\kappa}$ under the 
form of a gradient and time-derivative
expansion that corresponds to an expansion of all correlation functions in terms of their frequencies and momenta.

The {\it Ansatz}  for $\Gamma_\kappa$ at order one  in  time and  two in space derivatives  writes \cite{canet07}
\begin{equation}
\Gamma_\kappa[\Psi]=\int_{\vx,t} \,
 X_\kappa(\psi)\left(\tp\,\p_t \psi -\tp^2\right)+ \tp\,\left(U_\kappa'(\psi)- 
Z_\kappa(\psi)\,\nabla^2\,\psi -\frac{1}{2}\,\p_\psi Z_\kappa(\psi) (\nabla \psi)^2 \right).
\label{anz}
\end{equation}
and the initial conditions of the flow are $U_\Lambda'=V'$, $Z_\Lambda=1=X_\Lambda$.
Let us briefly justify this form: the time-reversal symmetry  requires  the \anz 
to be invariant  under the transformation  (\ref{trs}). It first implies that
 the $\tp^2$ and $\tp \p_t \psi$
 terms renormalize in the same way. These terms hence bear in (\ref{anz}) the same coefficient $X_\kappa(\psi)$.
 It then implies that the term linear in $\tp$ is invariant on its own, which imposes that it derives from a functional ${\cal H}_\kappa(\psi)$. We naturally choose
 for this functional the standard \anz for the equilibrium Ising model ${\cal H}_\kappa = \int Z_\kappa(\psi)(\nabla \psi)^2/2 + U_\kappa(\psi)$ \cite{canet07}. 
Finally, note that in \cite{canet07}, only a field-independent running coefficient  $X_\kappa(\psi)\equiv X_\kappa$
 was considered (leading order for the critical exponent $z$), whereas we here allow for a field 
dependence of $X_\kappa(\psi)$.

The definitions of the functions involved in Eq. (\ref{anz}) are
\begin{equation}
\begin{array}{l c l}
U_\kappa'(\psi) &=& \displaystyle{\rm FT}\left( \frac{\delta}{\delta\tp({\bf x})}\Gamma_\kappa \Big|_{\Psi=(\psi,0)}\right)\Big|_{\nu=0,\vec{p}=0}\\
 Z_\kappa(\psi) &=& \displaystyle \left[ \p_{\vec{p}\,^2} {\rm FT}\left( \frac{\delta^2}{\delta\tp({\bf x}) \delta\psi({\bf y})}\Gamma_\kappa\Big|_{\Psi=(\psi,0)}\right)\right]\Big|_{\nu=0,\vec{p}=0}\\
 X_\kappa(\psi)&=&\displaystyle \left[ \p_{i\nu} {\rm FT}\left(\frac{\delta^2}{\delta\tp ({\bf x})\delta\psi({\bf y})}\Gamma_\kappa\Big|_{\Psi=(\psi,0)}\right)\right]\Big|_{\nu=0,\vec{p}=0}
\end{array}
\label{defxz}
\end{equation}
where $\psi$ is a constant (in time and space) and where FT(.) means the Fourier transform. By translational   invariance  the momentum and frequency are vanishing in the first line and the second last two lines only depend 
on one momentum $\vec p$ and one frequency $\nu$ (see conventions of Eq. ({\ref{inv-trans}) -- the trivial $2\pi$ and $\delta(.)$ factors are not made explicit).} In the spirit of the derivative expansion, the renormalization functions are computed at vanishing external momentum and frequency since within this approximation only the small momentum and frequency sector is correctly described.

Their flow follows from the flow of the one- and two-point functions, derived from Eq. (\ref{dkgam}). For  $U_\kappa'$ for instance, it is given by 
\begin{equation}
\begin{array}{l c l}
 \p_\kappa U_\kappa'(\psi) &=& \displaystyle {\rm FT} \left( \left[\frac{\delta}{\delta\tp}\p_\kappa\Gamma_\kappa\right]\Big|_{\Psi=(\psi,0)}\right) \\
      &=&\displaystyle {\rm FT} \left(\frac 1 2 \tilde\p_{ \kappa}  \;\;{\rm Tr} \Big[ \displaystyle \int_{\{t_i,\vx_i\}} \!\; 
\Gamma_{\kappa,\tp}^{(3)}\; .\; G_\kappa \; \Big]\!\Big|_{\Psi=(\psi,0)} \right)
\label{dkupn}
\end{array}
\end{equation}
where the 3-point function is  put in the $2\times2$ matrix form $\Gamma_{\kappa,\tp}^{(3)} = \p\Gamma_\kappa^{(2)}/\p \tp$
and where $\tilde\p_{ \kappa} \equiv \p_\kappa R_\kappa \p /\p R_\kappa$. 
Taking the appropriate functional derivatives of (\ref{anz}) and evaluating the result at the uniform and stationary field 
configuration $\Psi(t,\vx) = (\psi,0)$ one finds in Fourier space (see Appendix \ref{appendixA} for notations):
\begin{equation}
\begin{array}{l c l}
 \Gamma_{\kappa,\tp}^{(3)}(\{\omega_i,\vq_i\};\psi)=  \left[
\begin{array}{c c}
  U_\kappa^{(3)} + Z'_\kappa(\vq_1^{\;2}+\vq_2^{\;2}+\vq_1.\vq_2) +i \omega_1 X'_\kappa & -2 X'_\kappa \\
 -2 X'_\kappa                                                                                                                                                & 0
\end{array}\right].
\end{array}
\label{gamma3}
\end{equation}
 The  propagator $G_\kappa$ in Eq. (\ref{dkupn}) is obtained by
 inverting the $2\times2$ matrix  $\left(\Gamma_\kappa^{(2)} + R_\kappa \right)$  evaluated in $\Psi(t,\vx) = (\psi,0)$  (or equivalently from the general expression (\ref{propagapp1},\ref{propagapp2}) of Appendix \ref{appendixA}). 
We find:
\begin{equation}
 G_\kappa(\omega,\vq;\psi) = \frac{1}{h(\vq)^2+(X_\kappa(\psi) \omega)^2} 
\left[
\begin{array}{c c}
 2 X_\kappa(\psi)                                                 &     h(\vq) +i \omega X_\kappa(\psi)  \\
 h(\vq) -i\omega X_\kappa(\psi)           &     0
\end{array}\right]
\label{prop}
\end{equation}
with $h(\vq) = Z_\kappa(\psi) \vq^{\;2} + R_\kappa(\vq^{\;2})+ U_\kappa''(\psi)$.
More precisely, implementing the regularization advocated in  Section \ref{ito}, the off-diagonal 
terms of this propagator should be replaced by 
\begin{equation}
 \displaystyle \frac{h(\vq) \pm i\omega X_\kappa(\psi) }{h(\vq)^2+(X_\kappa(\psi) \omega)^2}\; e^{\pm i\epsilon \omega}.  
\label{propeps}
\end{equation}
After Fourier transforming Eq. (\ref{dkupn}) and inserting the expressions (\ref{gamma3}) and (\ref{prop}), 
 the matrix product, trace and integral over the internal frequency $\omega$ are straightforward. The resulting flow equation for  
 $U_\kappa'$ is  expressed in term of a single integral over the internal momentum $\vq$. 
The flows of $Z_\kappa$ and $X_\kappa$ are computed using Eq. (\ref{dkgam2ito}), inserting the  expressions for $G_\kappa$, $\Gamma_\kappa^{(3)}$ and $\Gamma_\kappa^{(4)}$ (see Appendix \ref{appendixA}) and taking the appropriate derivatives  with respect to the external momentum and frequency respectively as in (\ref{defxz}).
Once again,  the  internal frequency integral can be calculated analytically, such that the flow equations  for $Z_\kappa$ and $X_\kappa$ involve each only one remaining  integral over the internal momentum $\vq$.

 Since we are interested in the scale invariant (fixed point) regime, we introduce dimensionless field and renormalization functions 
\begin{equation}
\begin{array}{l c l}
 \hr &=& \kappa^{(2-d)/2}\bZ_\kappa^{1/2} \psi\\
\hu(\hr) &=&\kappa^{-d} U_\kappa(\psi)\\
 \hz(\hr) &=& \bZ_\kappa^{-1} Z_\kappa(\psi)\\
\hx(\hr) &=& \bX_\kappa^{-1} X_\kappa(\psi)\\
\end{array}
\label{de-dim}
\end{equation}
where the running coefficients  $\bZ_\kappa\equiv Z_\kappa(\psi_0)$ and $\bX_\kappa\equiv X_\kappa(\psi_0)$
are defined at a fixed normalization point $\psi_0$.
In the critical regime, these running coefficients are expected to behave as power laws 
$\bZ_\kappa \sim \kappa^{-\etaz(\kappa)}$ and $\bX_\kappa \sim \kappa^{-\etax(\kappa)}$ with 
$\etaz(\kappa) = -\kappa \p_\kappa \ln Z_\kappa$ and similarly
  for $\etax(\kappa)$. The critical exponents $\eta$ and $z$ can  be expressed in terms of the fixed point values
of $\etaz(\kappa)$ and  $\etax(\kappa)$  as $\eta \equiv \etaz^*$  and $z\equiv 2-\etaz^* +\etax^*$. 

As already mentioned, the three flow equations for $U_\kappa', Z_\kappa$ and $X_\kappa$
 involve each a single integral over a momentum variable $\vq$.
 It is convenient to further introduce the dimensionless square internal momentum $y=q^2/\kappa^2$ and the dimensionless 
cutoff function $r(y) = (\bZ_\kappa q^2)^{-1} R_\kappa(q^2/\kappa^2)$. 
The flow equations for the dimensionless renormalization functions $\hu', \hz$ and $ \hx$ are the sum of  a dimensional
part that comes from the change of variables (\ref{de-dim}) and of a dynamical part that comes from the integration
of the rapid modes (previous calculations). The dimensional parts for $\hu', \hz$ and $ \hx$ are respectively:
\begin{equation}
\begin{array}{lll}
\p_s \hu'\Big|_{\rm dim}&= & \frac 1 2 \left( -\left(d+2 - \etaz\right)\hu'
      + \left( -2 + d + \etaz \right) \hr \,\hu''\right)\\
\p_s \hz\Big|_{\rm dim}&= & \etaz \hz  +\frac 1 2 \left( -2 + d + \etaz \right) \hr \,\hz' \\
\p_s \hx\Big|_{\rm dim}&= & \etax \hx  +\frac 1 2 \left( -2 + d + \etaz \right) \hr \,\hx'
\end{array}
\label{dsudim}
\end{equation}
and for the dynamical parts we find:
\begin{equation}
\begin{array}{lll}
\p_s \hu'\Big|_{\rm dyn}&= &-\displaystyle \frac{v_d}{2} \int d y\,y^{\frac d 2 -1} \frac{f s}{h^2} \\
\p_s \hz\Big|_{\rm dyn}&= & \displaystyle v_d \int dy \,y^{\frac d 2 -1} \frac{s }{h^2}\Big[
 \frac{f^2}{h^2}\left(- h' + \frac 4 d \frac{y h'^2}{h} -  \frac 2 d y h''\right)\\
 & &+ \displaystyle 2 \hz'\frac{f}{h}\left(1-\frac 2 d \frac{y h'}{h} \right) + \frac 1 d (\hz')^2  \frac{y}{h} - \frac{\hz''}{2} \Big]\\
 \p_s \hx\Big|_{\rm dyn} &= &\displaystyle  v_d \int dy  \,y^{\frac d 2 -1}\frac{s }{h^2}\left(\frac 3 4 
\frac{f^2}{h^2}-2 \hx' \frac{f}{h} + \frac 1 2 \hx'' \right)
\label{dsu}
\end{array}
\end{equation}
with $\p_s\equiv \kappa \p_\kappa$, $v_d^{-1}= 2^{d}\pi^{d/2}\Gamma(d/2)$, $h(y,\psi)=y(\hz(\psi)+r(y))+ \hu''(\psi)$, 
 $f(y,\psi) = y\hz'(\psi) +\hu'''(\psi)$ and  $s(y) = -\etaz r(y) -2 y r'(y)$.  
As expected,   $\hx$ does not contribute to  $\p_s \hu'$ and 
 $\p_s \hz$ which are the standard equilibrium equations of the Ising model at second order in the 
derivative expansion \cite{berges02,canet03a}. 
The numerical study of these equations  in dimensions two and three, 
which does not present any serious difficulty, will appear elsewhere.

\section{Supersymmetry }
\label{supersymmetry}

It is well-known that the field theory associated with the Langevin equation of a 
model that relaxes towards thermodynamic equilibrium is endowed with a supersymmetry. 
Here we show that the superfield formalism that 
follows from this property leads to a rather simple NPRG formalism and, 
most importantly for us, is free 
from any ambiguity coming from the continuous time-limit. 
It is therefore a benchmark for testing our
regularization prescription in It$\bar{\rm o}$'s discretization.

\subsection{Time reversibility and supersymmetry of the action}

 We here follow the first route sketched out in Section \ref{janssen} to treat the determinant  (\ref{jacobian}) which consists in
 rewriting it  using Grassmann variables $\eta$, $\bar\eta$ as in (\ref{jacsusy}). We here consider that $N(\phi)\equiv 1$.
Once the  Jacobian is exponentiated, one can finally integrate in (\ref{average-jacobian}) 
over the Gaussian noise distribution and deduce
the generating functional of correlation and response functions
\begin{equation}
{\cal Z}[j,\tilde{j}]=\int \cD\phi\, \cD [i\tphi]\,\cD\eta\, \cD\bar\eta\, e^{-{\cal S}[\phi,\tphi,\eta,\bar\eta]+\int_{t,\vx}j\phi+\tilde{j}\tphi}
\label{zns}
\end{equation}
with 
\begin{equation}
{\cal S}[\phi,\tphi,\eta,\bar\eta] =  \int_{t,\vx} \left\{\tphi\left(\partial_t\phi+ \coefu F(\phi)\right) - 
\coefd \tphi^2  -\eta\left(\partial_t +\coefu \frac{\delta F(\phi)}{\delta \phi}\right)\bar\eta\right\} .
\label{action}
\end{equation}
If $F$ derives from a Hamiltonian:
\begin{equation}
F(\phi(t,\vx))=\frac{\delta H(\phi)}{\delta \phi(t,\vx)}
\end{equation}
then the Langevin equation (\ref{langevin}) (with $N=1$) corresponds to a dynamics
that leads at long time to thermodynamic equilibrium. In this case, 
 the noise strength is related to the relaxation rate by Einstein relation 
 $\coefu=\coefd$ (where $k_B T$ is set to unity), and
 this coefficient  can  be conveniently scaled away. 
This equilibrium property implies that the action ${\cal S}$  possesses a supersymmetry, and it
admits  a compact form in the superspace in term of a (bosonic) superfield \cite{zinn}
\begin{equation}
\Phi(t,\vx,\btheta,\theta) \equiv \Phi(\varpi) =\phi(t,\vx) + \eta(t,\vx)\btheta  + 
\theta \bar\eta(t,\vx)+  \theta\btheta\tilde\phi(t,\vx)
\label{superfield}
\end{equation}
with $\varpi = (t,\vx,\btheta,\theta)$ and
where $\theta$ and $\btheta$ are two anticommuting Grassmann variables 
\begin{equation}
\{\theta,\btheta\} = \theta^2= \btheta^2 = 0.
\end{equation}
The integrals over these variables are defined as
\begin{equation}
\displaystyle\int  d\theta = \int d\btheta = 0 \hbox{\hspace{1cm}}\int  d\theta\, \theta = \int d\btheta\, \btheta = 1.
\end{equation}
 One introduces the differential operators
\begin{equation}
D = \coef \p_\theta -\btheta \p_t \hbox{,\hspace{1cm}} \bD =  \p_{\btheta}\hbox{,\hspace{0.5cm} with \hspace{0.5cm}}\{D,\bD\} =-\p_t.
\end{equation}
such that the generating functional takes the simple form
\begin{equation}
\label{z-susy}
{\cal Z}[J]=\int \cD\Phi\, e^{-{\cal S}[\Phi]+\int_{\varpi} J \Phi}
\end{equation}
where $\int_{\varpi}\equiv\int  d^d \vx \;d t \; d\btheta \;d\theta$, with the action
\begin{equation}
{\cal S}[\Phi] = \int_{\varpi} \bD \Phi D \Phi + \coef H(\Phi)
\label{actionsusy}
\end{equation}
and the supersource $ J = \tJ +\btheta \gamma +\bar{\gamma} \theta + \theta\btheta j$.
 The generators of the supersymmetry trnasformations are
\begin{equation}
Q = \p_{\theta} \hbox{\hspace{1cm}} \bQ = \coef \p_{\btheta} +\theta \p_t \hbox{\hspace{0.5cm}with\hspace{0.5cm}}\{Q,\bQ\} =\p_t.
\end{equation}
One can check that the action (\ref{actionsusy}) is invariant under the infinitesimal transformations
$\delta \Phi = \epsilon Q \Phi$ and  $\delta \Phi = \epsilon \bQ \Phi$. The operators 
$D$ and $\bD$ are hence covariant derivatives for the supersymmetry \cite{zinn}.

The correlation function between two superfields ${\cal C}(\varpi_1,\varpi_2) = \la \Phi(\varpi_1)\Phi(\varpi_2)\ra$ 
encodes all two-point correlation and response functions, and
the supersymmetry encompasses all the dynamical symmetries.  Indeed, the correlation function ${\cal C}$
 then vanishes under the action of each of the three generators $Q_{1 2}$, $\{Q,\bQ\}_{1 2}$ and 
$\bQ_{1 2}$ (where $Q_{1 2}\equiv Q_1+Q_2 = \p_{\theta_1} + \p_{\theta_2}$ and similarly for the others),  which implies respectively 
causality, time-translational invariance and time-reversal symmetry, and yields in turn the fluctuation-dissipation theorem.
The supersymmetric formalism thus provides an elegant and powerful framework to treat equilibrium dynamics.

 Let us emphasize that reaction-diffusion systems do not, in general, 
show time-reversal symmetry. This is particularly spectacular for phase transitions to
absorbing states. The directed percolation action (\ref{action-dirperco}) 
is invariant  under the  `rapidity symmetry' $\phi(t)\to -\phi^*(-t)$, $\phi^*(t)\to -\phi(-t)$
(which is a characteristic feature of the models belonging to this universality class), 
but it is not invariant under
the transformation  (\ref{trs}) and thus does not show time-reversal symmetry. 
For this reason, the field theory of this model cannot be recast  into the 
superfield formalism with a simple action such as (\ref{actionsusy}).

\subsection{The NPRG in the supersymmetric formalism}

It is almost straightforward to render supersymmetric the NPRG formalism.
Here again a scale dependent generating functional ${\cal Z}_\kappa[J]$ is built from  ${\cal Z}[J]$ defined in
 Eq. (\ref{z-susy})  by
adding a quadratic cutoff term 
 \begin{equation}
{\cal Z}_\kappa[J]=\int \cD\Phi\, e^{-{\cal S}[\Phi]-\Delta{\cal S}_\kappa[\Phi] +\int_{\varpi} J\Phi}
\label{Z}
\end{equation}
where $\Phi$ is the superfield (\ref{superfield}). The cutoff-term now writes 
\begin{equation}
\Delta{\cal S}_\kappa[\Phi] = \frac{1}{2}\int_{\varpi,\varpi'} \Phi(\varpi) R_\kappa( \varpi- \varpi') \Phi( \varpi'),
\label{deltask}
\end{equation}
where $R_\kappa( \varpi- \varpi') =R_\kappa({\bf x}-{\bf x'})\delta(\bar\theta-\bar\theta') \delta(\theta-\theta')$ and where
 the Dirac function for a Grassmann variable is simply the identity: $\delta(\theta)=\theta$.
The functional $\Gamma_\kappa$ is defined as the (modified) Legendre transform of   
$\log {\cal Z}_\kappa[J]$:
\begin{equation}
\Gamma_\kappa[\Psi] +\log {\cal Z}_\kappa[J]= \int_\varpi J \Psi - \Delta {\cal S}_\kappa[\Psi].
\label{legendre}
\end{equation}
with $\Psi = \la \Phi\ra$.
The exact flow equation for $\Gamma_\kappa$ is formally identical to the one for a scalar field theory:
\begin{equation}
\partial_\kappa \Gamma_\kappa  = 
\displaystyle \frac{1}{2}\, \int_{\varpi,\varpi'}\! \partial_\kappa R_\kappa(\varpi -\varpi') \; G_\kappa[\varpi,\varpi';\Psi]
\label{dkgamsusy}
\end{equation}
with $G_\kappa \equiv\left(\Gamma_\kappa^{(2)}+R_\kappa\right)^{-1}$  and   
$\Gamma_\kappa^{(n)}$  the $n$-th functional derivative of $\Gamma_\kappa$ with respect to the superfield
 $\Psi(\varpi)$ (see Appendix \ref{appendixsusy}). The whole formalism is thus extremely 
close to the one for equilibrium theories, the
only difference being the Grassmann dimensions that encompass  
the out-of-equilibrium aspect of the model.
Let us now show on the example of Model A how this formalism can be used 
in practice within the derivative expansion.

\subsection{Model A within the supersymmetric NPRG}

The supersymmetric action of Model A is given by Eqs. (\ref{actionsusy},\ref{H-ising}). 
 As already stressed, the derivative expansion consists in constructing an {\it Ansatz} for $\Gamma_\kappa$ that  captures  the low momentum
and frequency sector of the model. In the supersymmetric formalism, it  simply consists in a series expansion of $\Gamma_\kappa$
in powers of $\nabla$ and of the covariant derivatives  $D$ and $\bD$  that is both supersymmetric and $Z_2$-invariant.
 At order two in derivatives, it writes:
\begin{equation}
\Gamma_\kappa = \int_\varpi  \left\{ X_\kappa(\Psi)\bD\Psi D \Psi + \frac{1}{2}\, Z_\kappa(\Psi)\, (\nabla \Psi)^2 + U_\kappa(\Psi)\right\}.
\label{anza}
\end{equation}
By integrating over the Grassmann variables in (\ref{anza}), one can check that the resulting \anz coincides with the non-supersymmetric \anz (\ref{anz}) with the same functions $U_\kappa$, $Z_\kappa$ and $X_\kappa$.
The flow equations for these functions are obtained
by taking the appropriate functional derivatives of the flow equation (\ref{dkgamsusy}).
The effective potential is defined from $\Gamma_\kappa^{(1)}$ evaluated in a uniform and 
stationary superfield configuration $\Psi[\varpi] = \psi $ 
\begin{equation}
 \Gamma_\kappa^{(1)}\Big|_{\Psi=\psi} = U_\kappa'(\psi).
\end{equation}
Hence the flow equation of (the first derivative of) the potential is given by
\begin{equation}
\begin{array}{l  l}
 \p_\kappa U_\kappa'(\psi) &\hspace{-2mm}=\displaystyle \left[\frac{\delta}{\delta\Psi(\varpi)}\p_\kappa\Gamma_\kappa\right]\Big|_{\Psi=\psi}\\
      &\hspace{-2mm}= \displaystyle \frac 1 2 \tilde \p_\kappa\int_{\{\varpi_i\}}\!\Gamma_\kappa^{(3)}[\varpi,\varpi_1,\varpi_2] G_\kappa[\varpi_2,\varpi_1]  \!\Big|_{\Psi=\psi}.
\label{dkup}
\end{array}
\end{equation}
Note that the cutoff function is here chosen  time-independent (see Section \ref{ito}). As usual, the full propagator in Eq. (\ref{dkup})
is obtained by inverting $(\Gamma_\kappa^{(2)}+R_\kappa)$ evaluated in the constant superfield configuration $\Psi=\psi$
(see Appendix \ref{appendixsusy} for more details). We find in Fourier space:
\begin{equation}
\label{propag_super}
 G_\kappa(\omega,\vq,\Theta_1,\Theta_2) = a_1(\omega,\vq) \delta^2(\Theta_1-\Theta_2) + a_2(\omega,\vq) \bar\delta^2(\Theta_1,\Theta_2)+ a_3(\omega,\vq).
\end{equation}
with $\Theta\equiv(\theta,\btheta)$,
$\delta^2(\Theta_1-\Theta_2)\equiv\delta(\theta_1-\theta_2)\delta(\btheta_1-\btheta_2)$,      
$\bar\delta^2(\Theta_1,\Theta_2)\equiv (\theta_1-\theta_2)(\btheta_1+\btheta_2)$ and
\begin{equation}
\label{propag_super_bis}
\begin{array}{l c l}
 a_1(\omega,\vq) &=& h(\vq)/P(\vq,\omega,\psi)\\
a_2(\omega,\vq)&=&  - i \omega X_\kappa(\psi)/P(\vq,\omega,\psi) \\
a_3(\omega,\vq) &=& 2 X_\kappa(\psi)/P(\vq,\omega,\psi)
\end{array}
\end{equation}
where $P(\vq,\omega,\psi) = h^2(\vq)+(X_\kappa(\psi)\omega)^2 $ and $h(\vq) = Z_\kappa(\psi)\vec{q}^{\;2} + R_\kappa(\vec{q}^{\;2}) + U_\kappa''(\psi)$.
Note that within the supersymmetric formalism, the equivalent of the flow (\ref{flot-V-psi}), {\it i.e.}
 Eq. (\ref{dkgamsusy}) evaluated at $\Psi(\varpi)=\psi$,
 is  vanishing by virtue of the Grassmann directions. No regularization is needed, 
the Grassmann directions automatically encode the appropriate causality properties.

The $\Gamma_\kappa^{(3)}$ function evaluated in the same configuration is obtained by taking three functional derivatives of the \anz (\ref{anza}) 
and evaluating the result at $\Psi=\psi$. One finds in Fourier space, using the conventions of Eq. (\ref{inv-trans}), (see also 
Appendix \ref{appendixsusy}) :
\begin{equation}
\label{gamma3_super}
\begin{array}{l c l}
 \Gamma_\kappa^{(3)}({\bf q}_1,\Theta_1,{\bf q}_2,\Theta_2,\Theta_3) &=& \hspace{-0.2cm}
- 2 X_\kappa'(\psi) \left(\delta^2(\Theta_1-\Theta_2)+\delta^2(\Theta_1-\Theta_3)\right)\\
 &&  \hspace{-2.5cm}+X_\kappa'(\psi) \left( (\theta_1-\theta_3)(\btheta_1-\btheta_2) + (\theta_1-\theta_2)(\btheta_1-\btheta_3) \right)\\
&&  \hspace{-2.5cm}
+\Big(U^{(3)}_\kappa(\psi) -
Z_\kappa'(\psi)(\vq_1^{\;2} + \vq_2^{\;2} +\vq_1.\vq_2) \Big)\delta^2(\Theta_1-\Theta_2)\delta^2(\Theta_1-\Theta_3)\\
&&  \hspace{-2.5cm}+  2 i(\omega_1+\omega_2) X_\kappa'(\psi)(\theta_1-\theta_3)(\theta_1-\theta_2)(\btheta_1\btheta_2 + \btheta_3\btheta_2) + 
2\leftrightarrow3.
\end{array}
\end{equation}
where,  by translational invariance,  $\omega_3=-\omega_1-\omega_2$. Once Eq. (\ref{dkup}) is Fourier transformed in space and time, it is  straightforward to
insert in this equation the expressions (\ref{propag_super}),  (\ref{propag_super_bis}) and (\ref{gamma3_super}) 
and to perform the integrals over the Grassmann variables, which cancels out most of the terms. 
The non-vanishing contributions yield the flow of 
$ U_\kappa'(\psi) $, in which 
 the integral over the internal frequency $\omega$ can be achieved to  express this flow in term of the only  integral over  the internal momentum $\vec{q}$  that remains. The result for the dimensionless potential
is identical to the one obtained in the nonsuperymmetric formalism, Eq. (\ref{dsu}).

The flow equations for $Z_\kappa$ and $X_\kappa$ can be defined from the flow equation of $\Gamma_\kappa^{(2)}$
{  evaluated at the constant field configuration $\Psi[\varpi] = \psi$: 
\begin{equation}
\begin{array}{l c l}
  \p_\kappa Z_\kappa(\psi) &\hspace{-0.3cm}=& \hspace{-0.3cm}\displaystyle \left[\p_{\vec{p}\,^2} 
{\rm FT}\left( \p_\kappa\Gamma_\kappa^{(2)}(\varpi_1,\varpi_2) \Big|_{\Psi=\psi}\right)\right]\Big|_{\theta_i=\bar\theta_i=\nu=\vec{p}=0}\\
 \p_\kappa X_\kappa(\psi)&\hspace{-0.3cm}=& \hspace{-0.3cm}\displaystyle  \left[ \p_{i\nu} {\rm FT}\left(\p_\kappa\Gamma_\kappa^{(2)} (\varpi_1,\varpi_2)  \Big|_{\Psi=\psi}\right)\right]\Big|_{\theta_i=\bar\theta_i=\nu=\vec{p}=0}
\end{array}
\label{defxzsusy}
\end{equation}
 with the same conventions as in Eq. (\ref{defxz}) and with the flow of $\Gamma_\kappa^{(2)}$  given by}
\begin{equation}
 \partial_\kappa \Gamma_\kappa^{(2)} =  \int_{\{\varpi_i\}}\! 
\partial_\kappa R_\kappa \; G_\kappa \;\left( - \frac 1 2\Gamma_\kappa^{(4)} +\Gamma_\kappa^{(3)}  G_\kappa \Gamma_\kappa^{(3)} \!\right)\; G_\kappa,
\label{dkgam2}
\end{equation}
 where the arguments $(\{\varpi_i\};\psi)$ 
of the vertex functions and of the propagator are implicit.
The expression for
$\Gamma_\kappa^{(4)}$ is derived in Appendix \ref{appendixsusy}. Once again, after inserting (\ref{propag_super}),  
(\ref{propag_super_bis}) and the expression for
$\Gamma_\kappa^{(4)}$ into Eq. (\ref{defxzsusy}), after  performing the integrations 
over the Grassmann variables and over the internal frequency,  the flow equations
depend on a single momentum integral and the equations thus obtained are identical to 
those derived in the nonsupersymmetric case, Eqs. (\ref{dsu}).
This constitutes a nontrivial check  of the consistency of the regularization of the propagator proposed
 with It$\bar{\rm o}$'s discretization.

\section{Conclusion}
\label{conclu}

This paper was devoted to presenting in detail the NPRG formalism 
for \nequ stochastic models in 
statistical physics. Special emphasis has been dedicated to
the inspection of the consequences of It$\bar{\rm o}$'s choice 
of discretization within this  formalism. 
The main objectives were three-fold i) provide a reference including 
all the specificities of NPRG 
for \nequ systems  ii) set up the NPRG framework to 
 deal with  supersymmetry, iii) identify the relevant prescriptions in NPRG related to  
It$\bar{\rm o}$'s discretization and  device a systematic way to enforce these prescriptions.

Regarding ii) we  showed in Section \ref{supersymmetry} 
that the supersymmetric formalism can be simply 
incorporated within the NPRG framework, and we derived in this context the 
NPRG flow equations for Model A at second order in the derivative expansion.
(Of course, this supersymmetric version of NPRG is mostly useful 
 when time-reversal symmetry holds.)
As for iii), going back to the discrete version of the field theory ensueing 
from either a Janssen-de 
Dominicis or a Doi-Peliti  transformation,
 we analyzed the causality properties of the response functions and showed that 
these properties are preserved
under the NPRG flows.  We proposed a simple prescription -- 
 a regularization factor $e^{\pm i\epsilon\omega}$ 
in the propagator -- to enforce automatically the consequences of 
It$\bar{\rm o}$'s prescription within the NPRG flows
\footnote{Note that this prescription is very close in spirit to what is encountered within the  second-quantization formalism, for instance in solid state physics, 
 in the presence of a first order time-derivative.
Once the Hamiltonian is put in a normal ordered form,
partition functions can be written in term of a functional integral
 expressed in the coherent states basis. When taking the continuous-time limit,
the ambiguity of the value of $\theta(t=0)$ arises, as in the Doi-Peliti
formalism, and is solved in a way similar to the one advocated here, {\it via} a regularization factor
$\exp(\pm i\epsilon\omega)$ in the propagator.
See, {\it e.g.} G.D. Mahan, {\it Many-Particle Physics}, 2nd. edition, (Plenum, New York, 1990).}.
We illustrated this procedure on the non-supersymmetric 
calculation of the NPRG flow equations 
for Model A, which identify with the ones obtained in the supersymmetric version and thus constitute a non-trivial check of the consistency of the proposed prescription. We emphasize  that, of course,  the whole non-supersymmetric framework presented here  also applies to genuinely out-of-equilibrium problems.

To conclude, let us emphasize once more that, as the continuous-time field theories for  
\nequ models (derived either from a Janssen-de Dominicis or Doi-Peliti formalism) are ill-defined 
on their own, it is crucial to be aware of the implicit discretization chosen and of its implications. 
This applies in particular within the NPRG framework.
 This is the first time that an exhaustive account of the specificities of the NPRG methods for \nequ 
systems is provided, including a  detailed analysis of discretization problems in this context. This 
contribution, together with the regularization proposed, will serve as a reference for future NPRG calculations in \nequ models.\\
 
\noindent{\it Acknowledgement}\\

The authors are deeply indebted to N. Wschebor for fruitful discussions throughout the completion 
of this work, in particular for the sketch of an earlier proof of the $\tphi$ proportionality of 
\nequ effective actions and for his participation in the proofs of Appendix \ref{appendix0}.  
The authors also wish to thank N. Dupuis for useful discussions and A. Ran\c con for a careful 
reading of the manuscript.

\section{Appendix}

\subsection{The limit $\kappa\to\Lambda$: a proof that $\Gamma_\Lambda={\cal S}$}
\label{appendix0}
We  prove in the following that if the cutoff function $R_\kappa$  diverges at
$\kappa=\Lambda$, the effective average action $\Gamma_\Lambda$ becomes the bare action at this scale. 
For simplicity, we consider in the following only the case of a frequency-independent cutoff
function $R_\kappa$.
The proof is rather different for the field theories derived  from
a Langevin equation (Janssen-de Dominicis formalism) and from a master equation (Doi-Peliti formalism). We thus study both cases
separately.

\subsubsection{Master equations}
\ 

The Doi-Peliti formalism for master equations  naturally leads to field
theories written in terms of two fields $\phi$ and $\phi^*$ that are complex conjugate. It is 
convenient to write 
 \begin{equation}
\phi=\phi_1 +i \phi_2
\end{equation}
with $\phi_i$ real and to couple these fields to real sources $J_i$ in the following way:
 \begin{equation}
\label{sources-reelles-Doi}
{\cal Z}_\kappa[J_1,J_2] = \int {\cal D}\phi_1{\cal D}\phi_2 e^{-{\cal S}[\phi_1,i\phi_2] - \Delta{\cal S}_\kappa[\phi_1,i\phi_2] +\int J_1\phi_1+J_2 i \phi_2 } .
\end{equation}
Note that with this choice the fields  $\phi$ and $\phi^*$ are coupled to independent and real sources:
$2(J_1\phi_1+J_2 i \phi_2)=(J_1+J_2)\phi+(J_1-J_2)\phi^* $.
 From the  definition (\ref{sources-reelles-Doi}), it is straightforward to show that 
${\cal Z}_\kappa^*={\cal Z}_\kappa$ which is thus real. 
Note that this property follows from the fact that, although complex, the action ${\cal S}$ is a polynomial
in $\phi_1$ and $i\phi_2$ with real coefficients (the reaction rates). It is also straightforward to show that
\begin{equation}
\langle\phi_1\rangle^*= \langle\phi_1\rangle\ \ \ {\rm and}\ \ \ \langle i\phi_2\rangle^*= \langle i\phi_2\rangle.
\end{equation}
We define $\psi_1=\langle\phi_1\rangle$, $ \psi_2= \langle i\phi_2\rangle$ and $\psi=\langle\phi\rangle$,
 $\tilde\psi=\langle\phi^*\rangle$ that are all real and independent: $\psi=\psi_1+\psi_2$ and  $\tilde\psi=\psi_1-\psi_2$.
We choose in the following the natural and convenient cutoff term $ \Delta{\cal S}_\kappa[\phi,\phi^*]= \Delta{\cal S}_\kappa[\phi_1,i\phi_2]$ 
\begin{equation}
\label{cutoff-phi1-iphi2}
 \Delta{\cal S}_\kappa=\frac{1}{2}\int_{\bf q} R_\kappa(\vec{q}\,^2) \phi({\bf q}) \phi^*(-{\bf q})=
\frac{1}{2}\int_{\bf q} R_\kappa(\vec{q}\,^2)\left( \phi_1({\bf q})\phi_1(-{\bf q})+ \phi_2({\bf q})\phi_2(-{\bf q})  \right)
\end{equation}
which we moreover take frequency-independent.
We symbolically note it $1/2 \int R_\kappa (\phi_1^2 +\phi_2^2)$.
We then define the effective average  action as
\begin{equation}
\label{transfo-legendre-psi12}
\Gamma_\kappa[\psi_1,\psi_2]+W_\kappa[J_1,J_2]=\int_{\bf x}\left(J_1\psi_1+J_2\psi_2\right)
-\frac{1}{2}\int_{{\bf x},{\bf x'}}\tilde\psi({\bf x}) R_\kappa({\bf x}-{\bf x'})\psi({\bf x'}).
\end{equation}
with $W_\kappa=\log {\cal Z}_\kappa$.
Note that this definition leads to a minus sign in the last term when written
in terms of $\psi_1$ and $\psi_2$: $1/2 \int R_\kappa (\psi_1^2 -\psi_2^2)$.
It is now easy to prove from Eq. (\ref{transfo-legendre-psi12}) the following relation:
\begin{equation} 
\label{gamma-action}
e^{-\Gamma_\kappa[\psi_1,\psi_2]}=\int{\cal D}\phi_1{\cal D}\phi_2 e^{-{\cal S}[\phi_1+\psi_1,i\phi_2+\psi_2]-\Delta{\cal S}_\kappa[\phi_1,i\phi_2]+
\int \frac{\delta \Gamma_\kappa}{\delta\psi_1}\phi_1 +  \frac{\delta \Gamma_\kappa}{\delta\psi_2}i\phi_2} .
\end{equation}
The cutoff term $\Delta{\cal S}_\kappa[\phi_1,i\phi_2]$ in Eq. (\ref{cutoff-phi1-iphi2}) is positive and formally identical
to the equilibrium one for a $O(2)$ model. This implies that, as usual, if $R_\kappa$ becomes infinite when 
$\kappa\to\Lambda$ the only configuration of $\phi_1$ and $\phi_2$ that contributes to the functional integral in
Eq. (\ref{gamma-action}) is $\phi_1=\phi_2=0$. This leads to the expected result
 $\Gamma_\Lambda[\psi_1,\psi_2]={\cal S}[\phi_1=\psi_1,i\phi_2=\psi_2]$. Note that since the action ${\cal S}$ of
 a reaction-diffusion system is proportional to $\phi^*$, this result shows that $\Gamma_\Lambda$ is proportional 
to $\tilde\psi$.

\subsubsection{Langevin equations}
\

The situation is different  for Langevin equations because the fields $\phi$ and $\tilde\phi$ are respectively real and pure imaginary
(note that in Eq. (\ref{average-jacobian}) the integration over $\tilde\phi$ is performed on the imaginary axis).
In It$\bar{\rm o}$'s discretization  where Grassmann fields are not necessary, the action corresponding to a generic Langevin
equation is quadratic in $\tilde\phi$ (once the average over the noise distribution has been performed). It writes:
\begin{equation}
{\cal S}=\int \tilde\phi K(\phi) - \tilde\phi^2 N^2(\phi),
\end{equation}
where $K(\phi)\equiv \p_t\phi -  F(\phi)$.
The generating functional of correlation and response function writes
 \begin{equation}
\label{sources-reelles-Langevin}
{\cal Z}_\kappa[j,\tilde{j}] = \int {\cal D}\phi{\cal D}{i\tilde\phi}\, e^{-{\cal S}[\phi,\tilde\phi] - 
\Delta{\cal S}_\kappa[\phi,\tilde\phi] +\int j\phi+\tilde{j}\tilde\phi } 
\end{equation}
with both $j$ and $\tilde{j}$ real.
This expression is formally very close to Eq. (\ref{sources-reelles-Doi})
 up to the fact that the cutoff term writes 
\begin{equation}
\label{cutoff-langevin}
 \Delta{\cal S}_\kappa=\int_{\bf q} R_\kappa(\vec{q}\,^2) \phi({\bf q}) \tilde\phi(-{\bf q})
\end{equation}
and is thus also pure imaginary.  The definition of the  effective average  action is
\begin{equation}
\label{transfo-legendre-langevin}
\Gamma_\kappa[\psi,\tilde\psi]+W_\kappa[J,\tilde J]=\int_{\bf x}\left(j\psi+\tilde{j}\tilde\psi\right)
-\int_{{\bf x},{\bf x'}}\tilde\psi({\bf x}) R_\kappa({\bf x}-{\bf x'})\psi({\bf x'})
\end{equation}
where $\psi=\langle\phi\rangle$ and $\tilde\psi=\langle\tilde\phi\rangle$. Once again, upon
rewritting $\tilde\phi$ as $i\tilde{\tilde\phi}$ with  $\tilde{\tilde\phi}$ real, it is straightforward to show
that ${\cal Z}_\kappa[j,\tilde{j}]$,  $\psi$ and $\tilde\psi$ are real. Following the same steps as above one can 
derive the same relation as Eq. (\ref{gamma-action}) with $\phi_1$ and  $i\phi_2$ replaced respectively by 
 $\phi$ and $\tilde\phi$, and  $\psi_1$ and  $\psi_2$ replaced by  $\psi$ and $\tilde\psi$. Now, since  
 ${\cal S}$ is quadratic in $\tilde\phi$, it is possible to perform the functional integral over this field
(taking into account the fact that it is imaginary). We obtain:
\begin{equation}
\label{gamma-kappa-lang}
\begin{array}{ll}
 \displaystyle{e^{-\Gamma_\kappa}=\int{\cal D}\phi}&\hspace{-0.3cm}\displaystyle{\exp\left( -\int \frac{1}{4 N^2(\phi+\psi)}\left(K(\phi+\psi) - 
2 \tilde\psi N^2(\phi+\psi)+R_\kappa \phi -\frac{\delta \Gamma_\kappa}{\delta\tilde\psi}  \right)^2\right) }\\
&\\
&\hspace{-0.3cm}\times\;\displaystyle{\exp\left(\int\frac{\delta \Gamma_\kappa}{\delta\psi}\phi-\int \tilde\psi K(\phi+\psi) - 
\tilde\psi^2 N^2(\phi+\psi) \right)}.
\end{array}
\end{equation}
Once again, if $R_\kappa$ diverges when $\kappa\to\Lambda$ only the configuration $\phi=0$ contributes to the functional
integral above. By expanding $\Gamma_\Lambda$ as a power series in $\tilde\psi$:
\begin{equation}
\Gamma_\Lambda= \int A_0(\psi) + \tilde\psi A_1(\psi) + \tilde\psi^2 A_2(\psi)+\dots
\end{equation}
we find from Eq. (\ref{gamma-kappa-lang}) evaluated at $\phi=0$: $ A_0(\psi)=0$, $ A_1(\psi)=N^2(\psi)$ , 
$ A_2(\psi)=G(\psi)$ and all $ A_n(\psi)=0$ for $n>2$. We thus conclude that  
$\Gamma_\Lambda[\psi,\tilde\psi]={\cal S}[\phi=\psi,\tilde\phi=\tilde\psi]$  as expected. Note that this result proves that
$\Gamma_\Lambda[\psi,\tilde\psi]\propto \tilde\psi$.

\subsection{Formulas for the NPRG propagator and  vertex functions}
\label{appendixA}
In the nonsupersymmetric case, we only need space and time coordinates and we 
define ${\bf x}=(t,\vec{x})$ and ${\bf q}=(\omega,\vec{q})$. Our convention for the Fourier transform is
\begin{equation}
f({\bf x})=\int_{{\bf q}} f({\bf q}) e^{i(\vec{q}.\vx -\omega t)}
\end{equation}
where  the same symbol is (abusively) used for the function and its Fourier transform
and where
\begin{equation}
\int_{{\bf q}}=\int_{\omega,\vec{q}}=\int \frac{d\omega}{2\pi}\frac{d^dq}{(2\pi)^d}\ \ \ {\rm and}\ \ \ 
\int_{{\bf x}}=\int_{t,\vec{x}}=\int dt d^dx .
\end{equation}
For a translational invariant function $f({\bf x},{\bf y})$ which 
depends only on the difference $({\bf x}-{\bf y})$,  we define:
\begin{equation}
\label{inv-trans}
f({\bf q},{\bf q'})=(2\pi)^{d+1}\delta^{d+1}({\bf q}+{\bf q'})f({\bf q}),
\end{equation}
without again changing symbols for the function. The generalization to functions
of $n$ momenta is straightforward: 
$f({\bf q}_1,\dots, {\bf q}_n)=(2\pi)^{d+1}\delta^{d+1}(\sum_i{\bf q}_i)f({\bf q}_1,\dots, {\bf q}_{n-1})$.

The matrix of two-point connected correlation (and response) functions $\hat{W}^{(2)}_\kappa$ is by definition:
\begin{equation}
{W}^{(2)}_{\kappa;ij}[{\bf x},{\bf x'};J]=\frac{\delta^2 W_\kappa}{\delta J_i({\bf x})\delta J_j({\bf x'})}.
\end{equation}
The same definitions hold for the vertex functions ${\Gamma}^{(2)}_{\kappa;ij}[{\bf x},{\bf x'};\Psi]$ by replacing
${W}_\kappa$ by ${\Gamma}_\kappa$ and $J$ by $\Psi$. From now on, 
the $\kappa$ and $J$ (resp. $\Psi$) dependences of $W$ (resp. $\Gamma$) are not systematically indicated to alleviate the notations.
We introduce the alternative notation, {\it e.g.}
\begin{equation}
{W}^{(1,1)}_{\kappa}[{\bf x},{\bf x'};J]=\frac{\delta^2 W_\kappa}{\delta j({\bf x})\delta \tJ({\bf x'})}
\end{equation}
where the first (resp. second) exponent in $(1,1)$ refers to the number of derivatives
 with respect to $j$ (resp. $\tJ$). The relationship between
the two definitions of the connected two-point functions is therefore:
\begin{equation}
\hat{W}^{(2)}_{\kappa}[{\bf x},{\bf x'};J]=\left(
\displaystyle\begin{array}{ll}
\frac{\delta^2 W_\kappa}{\delta j({\bf x})\delta j({\bf x'})}  & \frac{\delta^2 W_\kappa}{\delta j({\bf x})\delta \tJ({\bf x'})}\\
\frac{\delta^2 W_\kappa}{\delta \tJ({\bf x})\delta j({\bf x'})}& \frac{\delta^2 W_\kappa}{\delta \tJ({\bf x})\delta \tJ({\bf x'})}
\end{array}
\right)=\left(
\begin{array}{ll}
W^{(2,0)}_\kappa[{\bf x},{\bf x'};J]    &      W^{(1,1)}_\kappa[{\bf x},{\bf x'};J]\\
 W^{(1,1)}_\kappa[{\bf x'},{\bf x};J]   &      W^{(0,2)}_\kappa[{\bf x},{\bf x'};J]
\end{array}
\right)
\end{equation}
that is 
\begin{equation}
\begin{array}{l}
{W}^{(2)}_{12}[{\bf x},{\bf x'}]={W}^{(1,1)}[{\bf x},{\bf x'}]={W}^{(2)}_{21}[{\bf x'},{\bf x}]\\
{W}^{(2)}_{11}[{\bf x},{\bf x'}]={W}^{(2,0)}[{\bf x},{\bf x'}]
\ \ {,}\ \ {W}^{(2)}_{22}[{\bf x},{\bf x'}]={W}^{(0,2)}[{\bf x},{\bf x'}].
\end{array}
\end{equation}
The same kind of relations exist between the two-point vertex functions $\Gamma^{(n,\tilde{n})}({\bf x},{\bf x'})$, $n+\tilde{n}=2$.
In Fourier space, this implies:
\begin{equation}
\label{W-hat}
\hat{W}^{(2)}_{\kappa}[{\bf q},{\bf q'};J]=\left(
\begin{array}{ll}
W^{(2,0)}_\kappa[{\bf q},{\bf q'};J]    &      W^{(1,1)}_\kappa[{\bf q},{\bf q'};J]\\
 W^{(1,1)}_\kappa[{\bf q'},{\bf q};J]   &      W^{(0,2)}_\kappa[{\bf q},{\bf q'};J]
\end{array}
\right).
\end{equation}
When these functions are evaluated in a uniform field configuration, this matrix becomes proportional
to $\delta^{d+1}({\bf q}+{\bf q'})$ and using the convention of Eq. (\ref{inv-trans}) yields:
\begin{equation}
\hat{W}^{(2)}_{\kappa}({\bf q};J)=\left(
\begin{array}{ll}
W^{(2,0)}_\kappa({\bf q})    &      W^{(1,1)}_\kappa({\bf q})\\
 W^{(1,1)}_\kappa({-\bf q})   &      W^{(0,2)}_\kappa({\bf q})
\end{array}
\right)\ \ \  {\rm for\  } J {\rm \  uniform}
\end{equation}
and similar relations for $\hat\Gamma^{(2)}_{\kappa}({\bf q};\Psi)$ when $\Psi$ is uniform.

The modified Legendre transform $\Gamma_\kappa$  (the ``effective average action'') is defined by
\begin{equation}
\Gamma_\kappa[\Psi]+W_\kappa [J]=\int_{\bf x}   { }^tJ.\Phi -\frac{1}{2}\int_{\bf x,x'} { }^t\Psi({\bf x}). 
\hat{R}_\kappa({\bf x-x'}).\Psi({\bf x}) .
\end{equation}
In Fourier space:
\begin{equation}
R_\kappa({\bf q})=\left(
\begin{array}{ll}
 0                                                       &    R_\kappa^{11}({\bf q})\\
 R_\kappa^{11}(-{\bf q})                &R_\kappa^{02}({\bf q})
\end{array}
\right)
\label{regul_app}
\end{equation}
where we have taken $R_\kappa^{20}({\bf q})=0$.
The true Legendre transform of ${W}_\kappa$ is therefore $\Gamma_\kappa[\Psi]+\frac{1}{2}\int { }^t\Psi. 
\hat{R}_\kappa.\Psi$. We thus find that the full field-dependent propagator is:
\begin{equation}
\label{G-hat}
\hat{G}_\kappa=(\hat{\Gamma}^{(2)}_\kappa+\hat{R}_\kappa)^{-1}=\hat{W}^{(2)}_\kappa
\end{equation}
since
\begin{equation}
\label{inverseW}
\int_{\bf y}\hat{W}^{(2)}_\kappa[{\bf x},{\bf y};J] .(\hat{\Gamma}^{(2)}_\kappa+\hat{R}_\kappa)[{\bf y},{\bf z};\Psi]=
\delta^{d+1}({\bf x}-{\bf z})\, {\rm I_2}
\end{equation}
where $I_2$ is the $2\times2$ unit matrix. In Fourier space this relation writes:
\begin{equation}
\int_{\bf q}\hat{W}^{(2)}_\kappa[{\bf p},{\bf q};J] .(\hat{\Gamma}^{(2)}_\kappa+\hat{R}_\kappa)[{-\bf q},{\bf p'};\Psi]=
(2\pi)^{d+1}\delta^{d+1}({\bf p}+{\bf p'})\, {\rm I_2}.
\end{equation}

The exact flow equation for $\Gamma_\kappa$ writes in Fourier space:
\begin{equation}
\partial_\kappa \Gamma_\kappa [\Psi]=\frac{1}{2} \Tr \int_{\bf q} \partial_\kappa\hat{R}_\kappa({\bf q}).
\hat{G}_\kappa[{\bf q},-{\bf q};\Psi]
\end{equation}
with $\hat{G}_\kappa$ defined by Eqs. (\ref{W-hat},\ref{G-hat},\ref{inverseW}).  
Note that this equation can be conveniently rewritten as
\begin{equation}
\partial_\kappa \Gamma_\kappa [\Psi]
 = \frac{1}{2}\,{\tilde\p_\kappa} \;\;{\rm Tr} \int_{{\bf x},{\bf x'}}\! \ln \left(\hat{\Gamma}^{(2)}_\kappa+\hat{R}_\kappa\right)[{{\bf x},{\bf x'}};\Psi]
\label{dkgamstand}
\end{equation}
where $\tilde\p_\kappa = \p_\kappa R_\kappa \p /\p_{R_\kappa}$, which is frequently used.

It is in general sufficient to consider the flows of the vertex functions evaluated
in a uniform field configuration $\Psi_u$ (for the calculations performed
within the derivative expansion for instance). 
In this case, the equations simplify since
the problem becomes translational invariant. The flow of the two-point functions in
the configuration $\Psi_u$ writes:
\begin{equation}
\label{flot-gam2-appendice}
\begin{array}{l}
\displaystyle {\partial_\kappa \Gamma^{(2)}_{\kappa,ij} ({\bf p};\Psi_u)
 = {\rm Tr} \int_{\bf q} \partial_\kappa\hat{R}({\bf q}).\hat{G}({\bf q}).  }\\
\displaystyle {\left( 
\hat{\Gamma}_i^{(3)}({\bf p},{\bf q},{-{\bf p}-{\bf q}}).\hat{G}({\bf p+q}).\hat{\Gamma}_j^{(3)}(-{\bf p},{\bf p+q},{-\bf q})
 -  \frac{1}{2}\,\hat{\Gamma}_{ij}^{(4)}({\bf p},{-\bf p},{\bf q},-{\bf q})\right).\hat{G}({\bf q})}
\end{array}
\end{equation}
where, in the uniform field configuration $\Psi_u$ and using the convention of Eq. (\ref{inv-trans})  :
\begin{equation}
\hat{G}_\kappa({\bf q},\Psi_u).(\hat{\Gamma}^{(2)}_\kappa({\bf q},\Psi_u)+\hat{R}_\kappa({\bf q}))=I_2
\end{equation}
and thus, omitting to write the $\Psi_u$ dependences:
\begin{equation}
\hat{G}_\kappa({\bf q})=\frac{1}{P({\bf q})}\left(
\begin{array}{ll}
-\Gamma_\kappa^{(0,2)}({\bf q})-R_\kappa^{02}({\bf q})  &  \Gamma_\kappa^{(1,1)}({\bf q})+R_\kappa^{11}({\bf q})\\
\Gamma_\kappa^{(1,1)}(-{\bf q})+R_\kappa^{11}(-{\bf q})  & -\Gamma_\kappa^{(2,0)}({\bf q})
\end{array}
\right)
\label{propagapp1}
\end{equation}
with
\begin{equation}
\begin{array}{ll}
P({\bf q})=&-\Gamma_\kappa^{(2,0)}({\bf q})(\Gamma_\kappa^{(0,2)}({\bf q})+R_\kappa^{02}({\bf q}))\\
&+(\Gamma_\kappa^{(1,1)}(-{\bf q})+R_\kappa^{11}(-{\bf q}))(\Gamma_\kappa^{(1,1)}({\bf q})+R_\kappa^{11}({\bf q})).
\end{array}
\label{propagapp2}
\end{equation}
The three-point functions appearing in Eq. (\ref{flot-gam2-appendice}) are defined by
\begin{equation}
\hat{\Gamma}_i^{(3)}[{\bf x},{\bf y},{\bf z};\Psi]=\frac{\delta\hat{\Gamma}^{(2)}[{\bf x},{\bf y};\Psi] }{\delta \Psi_i({\bf z})}
\end{equation}
and in (\ref{flot-gam2-appendice}), they are evaluated in a uniform field configuration and Fourier transformed.
The same kind of definition holds of course for the four-point functions. In a uniform field configuration, 
the precise relationship between the $\hat{\Gamma}_i^{(3)}$'s and the ${\Gamma}^{(n,\tilde{n})}$'s is:
\begin{equation}
\hat{\Gamma}_1^{(3)}({\bf p},{\bf q},-{\bf p}-{\bf q})=\left(
\begin{array}{ll}
\Gamma_\kappa^{(3,0)}({\bf p},{\bf q},-{\bf p}-{\bf q})  &  
\Gamma_\kappa^{(2,1)}({\bf p},{\bf q},-{\bf p}-{\bf q})\\
\Gamma_\kappa^{(2,1)}({\bf p},-{\bf p}-{\bf q},{\bf q})   &
\Gamma_\kappa^{(1,2)}({\bf p},{\bf q},-{\bf p}-{\bf q})
\end{array}
\right)
\end{equation}
and
\begin{equation}
\hat{\Gamma}_2^{(3)}({\bf p},{\bf q},-{\bf p}-{\bf q})=\left(
\begin{array}{ll}
\Gamma_\kappa^{(2,1)}({\bf q},-{\bf p}-{\bf q},{\bf p})  &  
\Gamma_\kappa^{(1,2)}({\bf q},{\bf p},-{\bf p}-{\bf q})\\
\Gamma_\kappa^{(1,2)}(-{\bf p}-{\bf q},{\bf p},{\bf q})   &
\Gamma_\kappa^{(0,3)}({\bf p},{\bf q},-{\bf p}-{\bf q})
\end{array}
\right) .
\end{equation}
and for the four-point functions:
\begin{equation}
\hat{\Gamma}_{12}^{(4)}({\bf p},-{\bf p},{\bf q},-{\bf q})=\left(
\begin{array}{ll}
\Gamma_\kappa^{(3,1)}({\bf p},{\bf q},-{\bf q},-{\bf p})  &  
\Gamma_\kappa^{(2,2)}({\bf p},{\bf q},-{\bf p},-{\bf q})\\
\Gamma_\kappa^{(2,2)}({\bf p},-{\bf q},-{\bf p},{\bf q})   &
\Gamma_\kappa^{(1,3)}({\bf p},{-\bf p},{\bf q},-{\bf q})
\end{array}
\right) 
\end{equation}
\begin{equation}
\hat{\Gamma}_{11}^{(4)}({\bf p},-{\bf p},{\bf q},-{\bf q})=\left(
\begin{array}{ll}
\Gamma_\kappa^{(4,0)}({\bf p},-{\bf p},{\bf q},-{\bf q})  &  
\Gamma_\kappa^{(3,1)}({\bf p},-{\bf p},{\bf q},-{\bf q})\\
\Gamma_\kappa^{(3,1)}({\bf p},-{\bf p},-{\bf q},{\bf q})   &
\Gamma_\kappa^{(2,2)}({\bf p},{-\bf p},{\bf q},-{\bf q})
\end{array}
\right) 
\end{equation}
\begin{equation}
\hat{\Gamma}_{22}^{(4)}({\bf p},-{\bf p},{\bf q},-{\bf q})=\left(
\begin{array}{ll}
\Gamma_\kappa^{(2,2)}({\bf q},-{\bf q},{\bf p},-{\bf p})  &  
\Gamma_\kappa^{(1,3)}({\bf q},{\bf p},-{\bf p},-{\bf q})\\
\Gamma_\kappa^{(1,3)}({-\bf q},{\bf p},-{\bf p},{\bf q})   &
\Gamma_\kappa^{(0,4)}({\bf p},{-\bf p},{\bf q},-{\bf q})
\end{array}
\right) .
\end{equation}

\subsection{Formulas for the supersymmetric formalism}
\label{appendixsusy}

In this appendix  the expressions of the supersymmetric propagator $G_\kappa[\varpi_1,\varpi_2;\Psi]$ 
and vertex functions 
 $\Gamma_\kappa^{(3)}$ and $\Gamma_\kappa^{(4)}$ are derived. 
For the propagator, one needs to determine the inverse of $\left(\Gamma_\kappa^{(2)}+R_\kappa\right)$.
Taking two successive functional derivatives of $\Gamma_\kappa$ defined by the \anz (\ref{anza}) 
with respect to $\Psi(\varpi_1)$ and $\Psi(\varpi_2)$ and evaluating the resulting expression at 
uniform and stationary superfield $\Psi(\varpi)\equiv \psi$ yields
\begin{equation}
\begin{array}{l c l}
 \left(\Gamma_\kappa^{(2)}+R_\kappa\right)(\varpi_1,\varpi_2) &=& \Big\{X_\kappa(\psi) (D \bD -\bD D)_{\varpi_1} - 
Z_\kappa(\psi)\,\nabla^2_{\vx_1} \\
 &+& R_\kappa(\vx_1-\vx_2)+ U_\kappa''(\psi)\Big\} \delta^{d+3}(\varpi_1-\varpi_2)
\end{array}
\end{equation}
with $\delta^{d+3}(\varpi_1-\varpi_2) \equiv \delta(t_1-t_2)\delta^d(\vx_1-\vx_2)\delta(\theta_1-\theta_2)  
\delta(\btheta_1-\btheta_2)$ and recalling that for Grassmann variables $\delta(\theta)= 
\theta$. It is convenient to Fourier transform this expression in space and time only (and not in the 
Grassmann directions since there is no  translational invariance in these directions). Using the convention of Eq.  
  (\ref{inv-trans}) -- that is without writing the ($2\pi$) and $\delta(.)$ factors resulting from  translational invariance in space and time -- 
we find:
\begin{equation}
\begin{array}{l c l}
 \left(\Gamma_\kappa^{(2)}+R_\kappa\right)({\bf q}_1,\Theta_1,\Theta_2) &=& \Big\{ \left(Z_\kappa(\psi)\vec{q}_1\hspace{0.05mm}^2+ 
R_\kappa(\vec{q}_1\hspace{0.05mm}^2) + U_\kappa''(\psi)\right)\delta^2(\Theta_1-\Theta_2)\\
&& + i\omega_1 X_\kappa(\psi)\bar\delta^2(\Theta_1,\Theta_2) - 2 X_\kappa(\psi)\Big\}
\end{array}
\end{equation}
with $\Theta\equiv(\theta,\btheta)$ and  
$\delta^2(\Theta_1-\Theta_2)\equiv\delta(\theta_1-\theta_2)\delta(\btheta_1-\btheta_2)$,      
$\bar\delta^2(\Theta_1,\Theta_2)\equiv (\theta_1-\theta_2)(\btheta_1+\btheta_2)$.
Let us now determine its inverse $G_\kappa$. As $\left(\Gamma_\kappa^{(2)}+R_\kappa\right)$ 
is diagonal in frequencies and momenta, so is $G_\kappa$.
The inversion relation  writes
\begin{equation}
 \int d\Theta_3  \;G_\kappa(\omega,\vq,\Theta_1,\Theta_3) \left(\Gamma_\kappa^{(2)}+R_\kappa\right)(\omega,\vq,\Theta_3,\Theta_2) = 
\delta^2(\Theta_1-\Theta_2).
\label{inverse}
\end{equation}
We search for the inverse  $G_\kappa$ of $\left(\Gamma_\kappa^{(2)}+R_\kappa\right)$
 assuming the same structure in the Grassmann directions, that is 
\begin{equation}
 G_\kappa(\omega,\vq,\Theta_1,\Theta_2) = a_1(\omega,\vq) \delta^2(\Theta_1-\Theta_2) + a_2(\omega,\vq) \bar\delta^2(\Theta_1,\Theta_2)+ a_3(\omega,\vq).
\end{equation}
The coefficients $a_i$ can then be calculated using Eq. (\ref{inverse}). One finds
\begin{equation}
\begin{array}{l c l}
 a_1(\omega,\vq) &=& h(\vq)/P(\vq,\omega,\psi)\\
a_2(\omega,\vq)&=&  - i \omega X_\kappa(\psi)/P(\vq,\omega,\psi) \\
a_3(\omega,\vq) &=& 2 X_\kappa(\psi)/P(\vq,\omega,\psi)
\end{array}
\end{equation}
where $P(\vq,\omega,\psi) = h^2(\vq)+(X_\kappa(\psi)\omega)^2 $ and $h(\vq) = Z_\kappa(\psi)q^2 + R_\kappa(q^2) + U_\kappa''(\psi)$.

Let us eventually give the expressions of the 3- and 4- point vertex functions. 
By taking an additional functional derivative of the \anz (\ref{anza}) and evaluating the result  at the uniform and stationary field configuration, one finds
\begin{equation}
\begin{array}{l c l}
 \Gamma_\kappa^{(3)}(\varpi_1,\varpi_2,\varpi_3) &=& U^{(3)}_\kappa(\psi) \delta^{d+3}(\varpi_1-\varpi_2) \delta^{d+3}(\varpi_1-\varpi_3)\\
 &-& Z_\kappa'(\psi)\Big[\delta^{d+3}(\varpi_1-\varpi_2)\nabla^2_{\vx_1}     \delta^{d+3}(\varpi_1-\varpi_3) \\
 &+&  \delta^{d+3}(\varpi_1-\varpi_3)\nabla^2_{\vx_1}     \delta^{d+3}(\varpi_1-\varpi_2) \\
 &+& \nabla_{\vx_1}\delta^{d+3}(\varpi_1-\varpi_2) \nabla_{\vx_1}\delta^{d+3}(\varpi_1-\varpi_3) \Big]\\
 &+& X_\kappa'(\psi)\Big[  \delta^{d+3}(\varpi_1-\varpi_2)(D\bD -\bD D)_{\varpi_1} \delta^{d+3}(\varpi_1-\varpi_3)\\ &+& \delta^{d+3}(\varpi_1-\varpi_3)(D\bD -\bD D)_{\varpi_1} \delta^{d+3}(\varpi_1-\varpi_2) \\
&+& D_{\varpi_1} \delta^{d+3}(\varpi_1-\varpi_2) \bD_{\varpi_1} \delta^{d+3}(\varpi_1-\varpi_3) \\
&+&
D_{\varpi_1} \delta^{d+3}(\varpi_1-\varpi_3) \bD_{\varpi_1} \delta^{d+3}(\varpi_1-\varpi_2) 
\Big]
\end{array}
\end{equation}
which in Fourier space writes
\begin{equation}
\begin{array}{l c l}
 \Gamma_\kappa^{(3)}({\bf q}_1,\Theta_1,{\bf q}_2,\Theta_2,\Theta_3) &=& \Bigg\{\Big(U^{(3)}_\kappa(\psi) -
Z_\kappa'(\psi)(\vq_1^{\;2} + \vq_2^{\;2} +\vq_1.\vq_2) \Big)\delta^2(\Theta_1-\Theta_2)\delta^2(\Theta_1-\Theta_3)\\
 &-& 2 X_\kappa'(\psi) \left(\delta^2(\Theta_1-\Theta_2)+\delta^2(\Theta_1-\Theta_3)\right)\\
 &+&  X_\kappa'(\psi) \left( (\theta_1-\theta_3)(\btheta_1-\btheta_2) + (\theta_1-\theta_2)(\btheta_1-\btheta_3) \right)\\
 &+&  2 i(\omega_1+\omega_2) X_\kappa'(\psi)(\theta_1-\theta_3)(\theta_1-\theta_2)(\btheta_1\btheta_2 + \btheta_3\btheta_2) + 2\leftrightarrow3.\Bigg\}.
\end{array}
\end{equation}
where,  by translational invariance,  $\omega_3=-\omega_1-\omega_2$.
The expression of the 4-point vertex function follows from the same procedure, which yields in Fourier space
\begin{equation}
\begin{array}{l c l}
\Gamma_\kappa^{(4)}({\bf q}_1,\Theta_1,{\bf q}_2,\Theta_2,{\bf q}_3,\Theta_3,\Theta_4) &=& \Bigg\{\left[  U^{(4)}_\kappa(\psi) +Z_\kappa''(\psi)\left(\vq_1^{\;2} +\vq_2^{\;2}+  \vq_3^{\;2}+ \vq_1.\vq_2 + \vq_1.\vq_3+\vq_2.\vq_3 \right)  \right]\\
&\times& \delta^2(\Theta_1-\Theta_2)\delta^2(\Theta_1-\Theta_3)\delta^2(\Theta_1-\Theta_4)\\
&-& 2 X_\kappa''(\psi) \Big[\delta^2(\Theta_1-\Theta_2)\delta^2(\Theta_1-\Theta_3)+ 3\leftrightarrow4+ 2\leftrightarrow3\Big]\\
&+&   X_\kappa''(\psi) \Big[\delta^2(\Theta_1-\Theta_2)(\theta_1-\theta_4)(\btheta_1-\btheta_3) +  3\leftrightarrow4   \\
&+&  \delta^2(\Theta_1-\Theta_3)(\theta_1-\theta_4)(\btheta_1-\btheta_2) +  2\leftrightarrow4 \\
 &+&  \delta^2(\Theta_1-\Theta_4)(\theta_1-\theta_2)(\btheta_1-\btheta_3) +  2\leftrightarrow3  \Big] \\
&-&   X_\kappa''(\psi) \Big[ i(\omega_1+\omega_2+\omega_3)(\theta_1-\theta_2)(\theta_1-\theta_3)(\theta_1-\theta_4)\\
 &\times& (\btheta_1\btheta_2 \btheta_3 + \btheta_2 \btheta_3\btheta_4)
  +   3\leftrightarrow4+ 2\leftrightarrow 4 \Big) \Bigg\},
\end{array}
\end{equation}
where again,  by translational invariance,  $\omega_4=-\omega_1-\omega_2-\omega_3$.


\end{document}